\documentclass[
  aps,
  prl,
  reprint,
  preprintnumbers,
  nofootinbib,nobibnotes]{revtex4-1}

\usepackage{amssymb}
\usepackage{amsmath}
\usepackage{graphicx,subfigure}
\usepackage{longtable}
\usepackage{verbatim}
\usepackage{amsfonts}
\usepackage{hyperref}
\usepackage{color}

\newcommand{\eq}[1]{Eq.~(\ref{#1})}
\newcommand{\eqsand}[2]{Eqs.~(\ref{#1}) and (\ref{#2})}

\newcommand{\Dbar}{\,\overline{\!D}}
\newcommand{\dd}{\ensuremath{D\!-\!\Dbar{}\,}}
\newcommand{\ddm}{\dd\ mixing}

\newcommand{\be}{\begin{equation}}
\newcommand{\ee}{\end{equation}}
\newcommand{\bea}{\begin{eqnarray}}
\newcommand{\eea}{\end{eqnarray}}
\newcommand{\beq}{\begin{equation}}
\newcommand{\eeq}{\end{equation}}
\def\beqa{\begin{eqnarray}}
  \def\eeqa{\end{eqnarray}}
\newcommand{\bv}{\left(\begin{array}{c}}
\newcommand{\ev}{\end{array}\right)}

\def\lsim{\mathrel{\rlap{\lower4pt\hbox{\hskip1pt$\sim$}}
    \raise1pt\hbox{$<$}}}	  
\def\gsim{\mathrel{\rlap{\lower4pt\hbox{\hskip1pt$\sim$}}
    \raise1pt\hbox{$>$}}}	  

\newcommand{\nn}{\nonumber}

\begin{document}

\preprint{TTP15-020}

\vspace*{-30mm}

\title{Sum Rules of Charm CP Asymmetries beyond the SU(3)$_F$ Limit}
\thanks{dedicated to the memory of Lincoln Wolfenstein (1923-2015)}
\author{Sarah M\"uller$^{\,a,b}$}
\email{sarah.mueller2@kit.edu}
\author{Ulrich Nierste$^{\,b}$}
\email{ulrich.nierste@kit.edu}
\author{Stefan Schacht$^{\,b}$}
\email{stefan.schacht@kit.edu}
\affiliation{
$^{\,a}$ Institut f\"ur Experimentelle Kernphysik, Karlsruher
  Institut f\"ur Technologie, 76021 Karlsruhe, Germany\\
$^{\,b}$ Institut f\"ur Theoretische Teilchenphysik, Karlsruher
  Institut f\"ur Technologie, 76128 Karlsruhe, Germany}

\vspace*{1cm}

\begin{abstract}
We find new sum rules between direct CP asymmetries in $D$
  meson decays with coefficients that can be determined from a
  global fit to branching ratio data. Our sum rules eliminate the
  penguin topologies $P$ and $PA$, which cannot be determined from
  branching ratios.  In this way we can make predictions about direct
CP asymmetries in the standard model without \emph{ad hoc} assumptions on
the sizes of penguin diagrams. We consistently include first-order
  SU(3)$_F$ breaking in the topological amplitudes extracted from the
  branching ratios. By confronting our sum rules with future precise data
  from LHCb and Belle~II one will identify or constrain new-physics
  contributions to $P$ or $PA$.
The first sum rule correlates the CP asymmetries $a_{CP}^{\mathrm{dir}}$
in $D^0\to K^+K^-$, $D^0\to \pi^+\pi^-$, and $D^0\to \pi^0\pi^0$.  We
study the region of the $a_{CP}^{\mathrm{dir}}(D^0\to
\pi^+\pi^-)$--$a_{CP}^{\mathrm{dir}} (D^0\to \pi^0\pi^0)$ plane allowed
by current data and find that our sum rule excludes more than half of
the allowed region at 95\% C.L. Our second sum rule correlates the direct
CP asymmetries in $D^+\rightarrow \bar{K}^0 K^+$, $D_s^+\rightarrow K^0
\pi^+$, and $D_s^+\rightarrow K^+\pi^0$.
\end{abstract}

\maketitle

\section{Introduction}
Decays of charmed mesons are currently the only way to probe flavor
violation in the up-quark sector. A major goal of experimental charm
physics is the discovery of CP violation in nonleptonic charm decays
(see,~e.g.,~\cite{Aaij:2011in, Collaboration:2012qw, Aaij:2014gsa,
  Nisar:2014fkc}). To this end, it is promising to study singly
Cabibbo-suppressed (SCS) decays~$d$ whose
direct CP asymmetries may be large enough to be detected in
the near future
\begin{align}
a_{CP}^{\mathrm{dir}}(d) &\equiv \frac{
	\vert \mathcal{A}^{\mathrm{SCS}}(d)\vert^2 - \vert 
\overline{\mathcal{A}}{}^{\mathrm{SCS}}({\overline d})
\vert^2
}{
	\vert \mathcal{A}^{\mathrm{SCS}}(d)\vert^2 + \vert 
\overline{\mathcal{A}}{}^{\mathrm{SCS}}({\overline d})\vert^2
}\,. \label{eq:defacp}
\end{align}
{Here,  $\mathcal{A}^{\mathrm{SCS}}(d)$ is a $D^0$ decay amplitude
and $ \overline{\mathcal{A}}{}^{\mathrm{SCS}}(\overline d)$ is the 
amplitude of the CP-conjugate $\overline{D}{}^0$ decay. We write} 
$\mathcal{A}^{\mathrm{SCS}}(d) = \mathcal{A}_{sd}^{\mathrm{SCS}}(d) +
\mathcal{A}_b^{\mathrm{SCS}}(d)$ with
\begin{align*}
\mathcal{A}_{sd}^{\mathrm{SCS}}(d) &= \lambda_{sd} \mathcal{A}_{sd}(d)
\,, 
&\mathcal{A}_b^{\mathrm{SCS}}(d) &=   -\frac{\lambda_b}{2} \, \mathcal{A}_b(d),
\end{align*}
and the shorthand notation $\lambda_q \equiv V_{cq}^* V_{uq}$ and
$\lambda_{sd} \equiv (\lambda_s-\lambda_d)/2$ for the elements of the
Cabibbo-Kobayashi-Maskawa matrix involved.  
{{The contribution of $\mathcal{A}_b(d)$ to branching ratios 
$\mathcal{B}(d)$ is negligible, which essentially entails 
$\mathcal{B}(d)\propto |\mathcal{A}_{sd} (d)|^2$.} 

Adopting the Particle Data Group convention (with $\lambda_s>0$) one may safely neglect
subleading powers of $\lambda_{b}/\lambda_{sd} \sim 10^{-3}$.
\eq{eq:defacp} reads
\begin{align}
  a_{CP}^{\mathrm{dir}}(d) &=
\frac{ \mathrm{Im}\,\lambda_b \; \mathrm{Im}\left[ e^{-i\delta(d)}
  \mathcal{A}_b(d) \right]}{ \vert
  \mathcal{A}^{\mathrm{SCS}}_{sd}(d)\vert } \label{eq:numACP-3}
\end{align}
in terms of the strong phase
$\delta(d)\equiv\mathrm{arg}\left[\mathcal{A}_{sd}(d)\right]$. The
smallness of $\mathrm{Im}\, \lambda_b$ renders
$a_{CP}^{\mathrm{dir}}(d)$ highly sensitive to physics beyond the
standard model (SM). To establish a ``smoking gun'' signal of new
physics one needs reliable SM predictions for
$a_{CP}^{\mathrm{dir}}(d)$ or at least robust theoretical upper bounds
on $|a_{CP}^{\mathrm{dir}}(d)|$ which cannot be exceeded within the SM.
The difficulty of such theory predictions can be witnessed from $\Delta
a_{CP}^{\mathrm{dir}}\equiv a_{CP}^{\mathrm{dir}}(D^0\rightarrow K^+K^-)
- a_{CP}^{\mathrm{dir}}(D^0\rightarrow\pi^+\pi^-)$: estimates vary
between $\mathcal{O}(0.01\%)$ \cite{Bigi:2011re}, $\mathcal{O}(0.1\%)$
\cite{Buccella:1994nf, Grossman:2006jg, Artuso:2008vf, Petrov:2010gy,
  Li:2012cfa}, $\sim -0.25\%$ \cite{Cheng:2012wr} and $\sim -0.4\%$
\cite{Brod:2011re}, not excluding an enhanced SM value between
$\sim-0.6\%$ and $\sim-0.8\%$ \cite{Brod:2011re,Pirtskhalava:2011va,
  Brod:2012ud,Feldmann:2012js,Hiller:2012xm,Franco:2012ck}.  There are
claims that CP-violating effects in charm physics can be
$\mathcal{O}(1\%)$ \cite{Golden:1989qx}. The situation is not any better
in CP violation induced by \ddm\ \cite{Bobrowski:2010xg}.  The key
problem is our lack of knowledge of the penguin amplitude entering
$\mathcal{A}_b(d)$ in \eq{eq:numACP-3} \cite{Feldmann:2012js,
  Brod:2011re, Brod:2012ud}.

All theoretical analyses of $D\to PP^\prime$ decays, where
$P$,$P^\prime$ denote pseudoscalar mesons, rely on the approximate
SU(3)$_F$ symmetry of the strong interaction \cite{Altarelli:1974sc,
  Kingsley:1975fe, Einhorn:1975fw, Voloshin:1975yx, Quigg:1979ic,
  Wang:1980ac, Chau:1982da, Golden:1989qx, Savage:1991wu, Chau:1991gx,
  Grossman:2006jg, Isidori:2011qw, Brod:2011re, Pirtskhalava:2011va,
  Cheng:2012wr,Bhattacharya:2012ah, Altmannshofer:2012ur,
  Feldmann:2012js, Brod:2012ud, Li:2012cfa, Franco:2012ck,
  Hiller:2012xm, Bhattacharya:2012pc, Bhattacharya:2012kq,
  Atwood:2012ac, Grossman:2012ry, Hiller:2012wf, Buccella:2013tya,
  Gronau:2013xba, Lenz:2013pwa, Gronau:2015rda, Muller:2015lua}.
{In analyses of branching ratios one can include first-order 
SU(3)$_F$ breaking
  \cite{Grossman:2006jg,Li:2012cfa, Cheng:2012wr,Brod:2011re,
    Pirtskhalava:2011va,Feldmann:2012js, Brod:2012ud,
    Hiller:2012xm,Franco:2012ck,Savage:1991wu,
    Chau:1991gx,Bhattacharya:2012ah, Bhattacharya:2012pc,
    Bhattacharya:2012kq,Grossman:2012ry,Buccella:2013tya,Gronau:2013xba,
    Gronau:2015rda,Muller:2015lua}.}  An intuitive way to exploit
SU(3)$_F$ relations involves topological amplitudes (pioneered in
Refs.~\cite{Wang:1980ac,Zeppenfeld:1980ex,Chau:1982da}) which
characterize the flavor flow in terms of tree ($T$), color-suppressed
tree ($C$), exchange ($E$), annihilation ($A$), penguin ($P_{d,s,b}$),
and penguin annihilation ($PA_{d,s,b}$) diagrams.  This method has been
extended to include linear SU(3)$_F$ breaking in applications to $B$
\cite{Gronau:1995hm} and $D$ \cite{Muller:2015lua} decays.
{{The first-order SU(3)$_F$-breaking corrections are parametrized by 
$C^{(1)}_{i}$,$A^{(1)}_{i},\ldots$, with $i=1,2,3$ labeling which $d$-quark line
is replaced by an $s$ line.}

With our inability to predict individual CP asymmetries it is natural to
study correlations among several asymmetries. There are two sum rules
which hold in the limit of exact SU(3)$_F$ symmetry
\cite{Grossman:2006jg,Grossman:2013lya,Hiller:2012xm}:
\begin{align}
  a_{CP}^{\mathrm{dir}}(D^0\rightarrow K^+K^-) +
  a_{CP}^{\mathrm{dir}}(D^0\rightarrow \pi^+\pi^-) &=
  0\,, \label{eq:su3limit1}\\ 
a_{CP}^{\mathrm{dir}}(D^+\rightarrow
  \bar{K}^0K^+) + a_{CP}^{\mathrm{dir}}(D_s^+\rightarrow K^0\pi^+) &=
  0\,. \label{eq:su3limit2}
\end{align}
Analyses of branching ratios permit the determination of $\vert
\mathcal{A}^{\mathrm{SCS}}_{sd}(d)\vert$ in \eq{eq:numACP-3} and,
through global fits, also to constrain the phase $\delta(d)$.  The
branching ratios of the decays entering
\eqsand{eq:su3limit1}{eq:su3limit2} exhibit sizable SU(3)$_F$ breaking
which limits the power of these SU(3)$_F$-limit sum rules to test the
SM.  In this Letter, we derive new sum rules which incorporate SU(3)$_F$
breaking in $\mathcal{A}^{\mathrm{SCS}}_{sd}(d)$ to linear order. To
this end, we use the result of our global fit to $D\to PP^\prime$
branching ratios in Ref.~\cite{Muller:2015lua} in two ways: On one hand,
we extract $\vert \mathcal{A}^{\mathrm{SCS}}_{sd}(d)\vert$ and
$\delta(d)$ for the decays of interest from the fit to find the
numerical relation between $a_{CP}^{\mathrm{dir}}(d)$ and
$\mathcal{A}_b(d)$ in \eq{eq:numACP-3}.  On the other hand, the same fit
also returns the values of the topological amplitudes entering the
fitted branching ratios. However, the desired $\mathcal{A}_b(d)$ of
individual decay modes involve additional topological amplitudes (new
combinations of $P_{d,s,b}$ and $PA_{d,s,b}$) which do not appear in the
branching ratios. Our sum rules are constructed in a way to eliminate
these unknowns. Unlike \eqsand{eq:su3limit1}{eq:su3limit2}, these new sum
rules use the SU(3)$_F$ limit for the eliminated $P_{d,s,b}$,
$PA_{d,s,b}$ in $\mathcal{A}_b(d)$ only, while consistently including
SU(3)$_F$ breaking in {the SM-dominated quantities} $T$, $A$, $C$,
$E$. {Since no relations among direct CP asymmetries hold to first
  order in SU(3)$_F$ breaking \cite{Grossman:2013lya}, this is the best
  which can be achieved.}

\section{CP Asymmetry Sum Rules}
Our analysis starts with the decomposition of $\mathcal{A}_{sd}(d)$ 
and $\mathcal{A}_b(d)$ in terms of topological amplitudes. 
As an example consider
\begin{align}    
  \mathcal{A}_{sd} (D^0\to \pi^+\pi^-)&
           =-T-E+P_{\mathrm{break}} \label{eq:ex0} \\
  \mathcal{A}_b (D^0\to \pi^+\pi^-)&=T+E+P+PA  \label{eq:ex1}
\end{align} 
with the SU(3)$_F$-breaking penguin topology
$P_{\mathrm{break}}\equiv P_s - P_d $ and $s,d$ denoting the quark
flavor in the penguin loop. $\mathcal{A}_b$ involves the new
penguin topologies $P\equiv P_d + P_s - 2 P_b$ and $PA\equiv PA_d + PA_s
- 2PA_b$, which do not appear in $ \mathcal{A}_{sd} (d)$ of any
  decay $d$ and, consequently, cannot be constrained by data on branching
  ratios. Next we combine \eqsand{eq:ex0}{eq:ex1} to eliminate the
  numerically large parameters $T$ and $E$ from $\mathcal{A}_b$
\begin{align}
 \mathcal{A}_b & (D^0\to \pi^+\pi^-)\nn\\  
&=-\mathcal{A}_{sd} (D^0\to \pi^+\pi^-)+
    P_{\mathrm{break}}+P+PA\,. 
\end{align}
{{Trading $T$ and $E$ for $\mathcal{A}_{sd}$ is possible for all
  decays $d$ considered in this Letter: With} 
$\mathcal{T}_i={C,P_{\rm    break}},\ldots$, {we write}
\begin{align}
\mathcal{A}_b (d) = {c_{sd}^d \mathcal{A}_{sd}(d) +}
\sum_i c_i^d \mathcal{T}_i \, , \label{eq:atop}
\end{align}
with $c_i^d$
specified in Table~\ref{tab:topoparametrization-subleading}.  Since
$e^{-i\delta(d)}\mathcal{A}_{sd}(d)=|\mathcal{A}_{sd}(d)|$ is real,
$c_{sd}^d \mathcal{A}_{sd}(d)$ does not contribute to
$a_{CP}^{\mathrm{dir}}(d)$ in \eq{eq:numACP-3}.
\begin{table*}
\begin{center}
\begin{tabular}{ccccccccccc}
  \hline \hline Decay amplitude $\mathcal{A}(d)$ & $A_{sd}(d)$ & $A_1^{(1)}$
  & $A_2^{(1)}$ & $A_3^{(1)}$ & $C+A$ & $C_3^{(1)}$ & 
 $P_{\mathrm{break}}$
  & $P{{+}2A}$ & $P{+PA}$ \\\hline
  $\mathcal{A}(D^0\rightarrow K^+ K^-)$ & 1 & 0 & 0 & 0 & 0 & 0 & $-1$ &
  {0} & 1 \\\hline
  $\mathcal{A}(D^0\rightarrow \pi^+ \pi^-)$ & $-1$ & 0 & 0 & 0 & 0 & 0 &
  1 & {0} & 1 \\\hline
  $\mathcal{A}(D^0 \rightarrow \pi^0 \pi^0)$ & $-1$ & 0 & 0 & 0 & 0 & 0
  & $-\frac{1}{\sqrt{2}}$ & {0} &
  $-\frac{1}{\sqrt{2}}$ \\\hline
  $\mathcal{A}(D^+ \rightarrow \bar{K}^0 K^+)$ & 1 & 0 & 0 & 2 & 0 & 0
  &$-1$ & 1 & 0 \\\hline
  $\mathcal{A}(D_s^+ \rightarrow K^0 \pi^+)$ & $-1$ & 2 & 2 & 0 & 0 & 0
  & 1 & 1 & 0 \\\hline
  $\mathcal{A}(D_s^+ \rightarrow K^+ \pi^0)$ & 1 & 0 & 0 & 0 &
  $\sqrt{2}$ & $\sqrt{2}$ & $\frac{1}{\sqrt{2}}$ & $-\frac{1}{\sqrt{2}}$
  & 0 \\\hline\hline
\end{tabular}
\caption{The coefficients {$c_i^d$}
of the topological decomposition of
  $\mathcal{A}_b(d)$ in \eq{eq:atop}. {$A_{1,2,3}^{(1)}$ and
    $C_3^{(1)}$ are first-order SU(3)$_F$-breaking corrections  to $A$
  and $C$, respectively, as defined in \cite{Muller:2015lua}.} 
\label{tab:topoparametrization-subleading}}
\end{center}
\end{table*}
\begin{table*}
\begin{center}
\begin{tabular}{ccc}
\hline \hline
Decay $d$ &  ${X(d)}$  & $\mathcal{S}(d)$ \\\hline
$D^0\rightarrow K^+ K^-$        & $e^{-i \delta\left( D^0\rightarrow K^+ K^-\right)} \left( - P_{\mathrm{break}} \right)$  & $ \left( P + PA \right)- e^{2 i \delta\left( D^0\rightarrow K^+ K^-\right) }\left( P + PA \right)^*$  \\\hline
$D^0\rightarrow \pi^+ \pi^- $   & $e^{-i \delta\left( D^0\rightarrow \pi^+ \pi^-\right)} \left( P_{\mathrm{break}} \right)$  & $\left( P + PA \right) -  e^{2 i \delta\left( D^0\rightarrow \pi^+ \pi^-\right) }   \left( P + PA \right)^* $  \\\hline
$D^0 \rightarrow \pi^0 \pi^0$   & $e^{-i \delta\left( D^0 \rightarrow \pi^0 \pi^0\right)} \left( -\frac{1}{\sqrt{2}} P_{\mathrm{break}}\right)$   & $\frac{1}{\sqrt{2}}\left(  - P - PA \right)-  e^{2 i \delta\left( D^0 \rightarrow \pi^0 \pi^0\right) }  \frac{1}{\sqrt{2}}\left(  - P - PA \right)^* $  \\\hline
$D^+ \rightarrow \bar{K}^0 K^+$ &  $e^{-i \delta\left( D^+ \rightarrow \bar{K}^0 K^+\right)} \left( -2 A^{\mathrm{fac}}_{ D^+ \rightarrow \bar{K}^0 K^+} + 2 \delta_A - P_{\mathrm{break}}\right)$  & $P  -   e^{2 i \delta\left( D^+ \rightarrow \bar{K}^0 K^+\right) }    P^* $  \\\hline
$D_s^+ \rightarrow  K^0 \pi^+$    & $e^{-i \delta\left( D_s^+ \rightarrow  K^0 \pi^+\right)} \left( 2 A^{\mathrm{fac}}_{ D_s^+ \rightarrow K^0 \pi^+} + 2 \delta_A + P_{\mathrm{break}} \right)  $   & $P - e^{2 i \delta\left( D_s^+ \rightarrow  K^0 \pi^+\right) } P^* $  \\\hline
$D_s^+ \rightarrow  K^+ \pi^0$    & $e^{-i \delta\left( D_s^+ \rightarrow  K^+ \pi^0\right)} \left( \sqrt{2} {C} + \sqrt{2} C_3^{(1)} + \frac{1}{\sqrt{2}} P_{\mathrm{break}}\right)$    & $\frac{1}{\sqrt{2}}\left( - P\right) -    e^{2 i \delta\left( D_s^+ \rightarrow  K^+ \pi^0\right) }  \frac{1}{\sqrt{2}}\left( - P\right)^* $  \\\hline\hline
\end{tabular}
\caption{Definitions of $X(d)$ and results for $\mathcal{S}(d)$ as used
  and defined in Eq.~(\ref{eq:defineS}) in case of the SU(3)$_F$ fit
  including $1/N_c$ {counting}.  For the sign conventions of
  $A^{\mathrm{fac}}(d)$ see Ref.~\cite{Muller:2015lua}. Note
  that {$A^{\mathrm{fac}}_{D^+
    \rightarrow \bar{K}^0 K^+} = 0$ by} isospin {symmetry}.
\label{tab:S}}
\end{center}
\end{table*}

Our strategy involves three steps: In step~1, we determine {all
  quantities entering $a_{CP}^{\mathrm{dir}}(d)$ in \eq{eq:numACP-3}
  except for $P$ and $PA$ from} a global fit to branching ratio data as
described in Ref.~\cite{Muller:2015lua}. {These fitted quantities are
  the topological amplitudes} {$A_i^{(1)}$}, $C$, $C_3^{(1)}$, {and}
$P_{\mathrm{break}}$ {entering \eq{eq:atop}} as well as
$|\mathcal{A}_{sd}(d)|$ and the phases $\delta(d)$.  {Note that
  $|\mathcal{A}_{sd}(d)|$ is trivially found from $\mathcal{B}(d)\propto
  |\mathcal{A}_{sd} (d)|^2$. As an important feature of charm physics,
  the large number of different branching fractions gives 
  useful information not only on the magnitudes of the topological
  amplitudes, but also on their phases (up to an overall unphysical
  phase). By plugging the results back into $\mathcal{A}_{sd}(d)$, we
  find $\delta(d)$ (up to discrete ambiguities).}

In step~2 we eliminate all hadronic parameters but $P$ and $PA$ from $
a_{CP}^{\mathrm{dir}}(d)$. To this end, we define
\begin{align}
\mathcal{S}(d) & \equiv 2i e^{i \delta(d)}
\left[\frac{\vert\mathcal{A}^{\mathrm{SCS}}_{sd}(d)\vert
    a_{CP}^{\mathrm{dir}}(d)}{ \mathrm{Im}\, \lambda_b}   -
  \mathrm{Im}\, X(d) \right] 
 \,, \label{eq:defineS}
\end{align}
with {$X(d)$} given in
Table~\ref{tab:S}. {$S(d)$ is calculated solely from experimental
input, as all ingredients of \eq{eq:defineS} are found from the global
fit to branching ratios. To relate $S(d)$ to theoretical quantities, we
use \eq{eq:numACP-3}}
\begin{align}
\mathcal{S}(d) &=  2i e^{i \delta(d)}  \mathrm{Im}
  \left[ e^{-i \delta(d)}\mathcal{A}_b(d)-X(d)	 \right]
 \,. \label{eq:defineS2}
\end{align}
For our example above {the subtraction term $-\mathrm{Im}\, X(d)$
removes $P_{\mathrm{break}}$} and  \eq{eq:defineS2} gives 
\begin{align}
& \mathcal{S} (D^0\to \pi^+\pi^-) \nn \\
&\; =  2i e^{i \delta(D^0\to \pi^+\pi^-)} 
     \mathrm{Im} \left[ e^{-i \delta(D^0\to \pi^+\pi^-)} (P+PA) \right]. 
\label{eq:ex}
\end{align} 
The right column of Table~\ref{tab:S} shows $\mathcal{S}(d)$ in terms of
$P$, $PA$, their complex conjugates {$P^*$, $PA^*$}, and $\delta(d)$
(which is determined from the fit) for all decays. (To relate \eq{eq:ex}
to the entry in Table~\ref{tab:S}, use $2 i \mathrm{Im} z = z- z^*$ for
$z= e^{-i \delta(D^0\to \pi^+\pi^-)} (P+PA)$.)

In step 3 {of our analysis} 
we construct {two} sum rules eliminating $P$, $PA$, $P^*$, and
$PA^*$, {each of which connects three $a_{CP}^{\mathrm{dir}}(d)$.}  
{Table~\ref{tab:S} reveals that the combined information 
from $\mathcal{S} (D^0\to \pi^+\pi^-)$ and $\mathcal{S} (D^0\to
\pi^0\pi^0)$ determines the complex quantity  $P+PA$. Any additional 
CP asymmetry depending on  $P+PA$ will then probe the
standard model; i.e.,\ with three CP asymmetries we can construct 
the desired sum rule eliminating $P+PA$:}\\ 
\textbf{CP Asymmetry Sum Rule 1:}
\begin{align}
&\frac{\mathcal{S}(D^0\rightarrow K^+K^-) - \mathcal{S}(D^0\rightarrow \pi^+\pi^- )}{e^{2 i \delta(D^0\rightarrow K^+K^-)} - e^{2 i \delta(D^0\rightarrow \pi^+\pi^-)}} - \nn\\
&\frac{\mathcal{S}(D^0\rightarrow K^+K^-) + \sqrt{2} \mathcal{S}(D^0\rightarrow \pi^0\pi^0 )}{e^{2 i \delta(D^0\rightarrow K^+K^-)} - e^{2 i \delta(D^0\rightarrow \pi^0\pi^0)}}
= 0 \,. \label{eq:sumrule1}
\end{align}
In this last step SU(3)$_F$ breaking in $P$, $PA$ is neglected. {The second sum rule, which
correlates three CP asymmetries depending on $P$, is}\\
\textbf{CP Asymmetry Sum Rule 2:}
\begin{align}
&\frac{\mathcal{S}(D^+\rightarrow \bar{K}^0 K^+) - \mathcal{S}(D_s^+\rightarrow K^0 \pi^+)}{e^{2i\delta( D^+\rightarrow \bar{K}^0 K^+) } - e^{2i\delta( D_s^+\rightarrow K^0 \pi^+) } } - \nn\\
&\frac{\mathcal{S}(D^+\rightarrow \bar{K}^0 K^+) + \sqrt{2} \mathcal{S}(D_s^+\rightarrow K^+\pi^0)}{ e^{2i\delta( D^+\rightarrow \bar{K}^0 K^+) } - e^{2i\delta(D_s^+\rightarrow K^+\pi^0) } }
= 0 \,. \label{eq:sumrule2}
\end{align}
{By inserting the expressions in the right column of
  Table~\ref{tab:S} one readily verifies
  \eqsand{eq:sumrule1}{eq:sumrule2}.}  If some of the phases in the
denominators of the sum rules are equal (covering the SU(3)$_F$ limit as
a special case) one finds: for $e^{2i\delta(D^0\rightarrow K^+K^-)} =
e^{2i\delta(D^0\rightarrow \pi^+\pi^-)} $ \eq{eq:sumrule1} collapses to
	\begin{align}
	\mathcal{S}(D^0\rightarrow K^+K^-) - \mathcal{S}(D^0\rightarrow \pi^+\pi^-) &= 0\,, \label{eq:specialFirst}
	\end{align}
while for $e^{2i\delta(D^0\rightarrow K^+K^-)} =
e^{2i\delta(D^0\rightarrow \pi^0\pi^0)} $ the sum rule becomes
	\begin{align}
	  \mathcal{S}(D^0\rightarrow K^+K^-) + \sqrt{2}
	  \mathcal{S}(D^0\rightarrow \pi^0\pi^0 ) &= 0\,.
	    \label{eq:specialMiddle}
	\end{align}
If all three phase factors are equal,
\eqsand{eq:specialFirst}{eq:specialMiddle} hold simultaneously.
The special cases of sum rule {2} are obtained from those
in Eqs.~(\ref{eq:specialFirst}) and (\ref{eq:specialMiddle}) by obvious
replacements.

When linking sum rule {2} to experimental quantities, we use
$a_{CP}^{\mathrm{dir}}( D^+\rightarrow K_S K^+ ) =
a_{CP}^{\mathrm{dir}}( D^+\rightarrow \bar{K}^0 K^+ )$ and
$a_{CP}^{\mathrm{dir}}( D_s^+\rightarrow K_S \pi^+) =
a_{CP}^{\mathrm{dir}}( D_s^+\rightarrow K^0\pi^+ )$, meaning that, in our
definition of $a_{CP}^{\mathrm{dir}}$ kaon CP violation is properly
subtracted \cite{Grossman:2011zk}.  The two sum rules probe the
$SU(3)_F$ limit in {$P$} and $P+PA$. If future experiments find
deviations of order $30\%$, one will ascribe those to $SU(3)_F$-breaking
hadronic effects. The smallness of $\mathrm{Im}\, \lambda_b$ makes the
sum rules highly sensitive to new physics, which may well violate the
sum rules at a far higher level.

\section{SM Prediction of CP Asymmetries}

\begin{table}
\begin{center}
\begin{tabular}{ccr}
\hline\hline
Observable	      &      Measurement &  References	\\\hline
$\Delta a_{CP}^{\mathrm{dir}}$	 & $-0.00253\pm 0.00104$ & \cite{Amhis:2012bh,Aubert:2007if,Aaij:2011in,Staric:2008rx, Collaboration:2012qw, Ko:2012px, LHCb:2013dka, Aaij:2013bra, Aaij:2014gsa} \\
$\Sigma a_{CP}^{\mathrm{dir}}$ & $-0.0011\pm 0.0026$ & $^\dagger$\cite{Ko:2012px,Aubert:2007if,Aaij:2011in,Aaltonen:2011se,Collaboration:2012qw}   \\
$a_{CP}^{\mathrm{dir}}(D^0 \rightarrow K_S K_S)$ & $-0.23\pm 0.19$ & \cite{Bonvicini:2000qm} \\
$a_{CP}^{\mathrm{dir}}(D^0 \rightarrow \pi^0 \pi^0)$   &  $-0.0004\pm 0.0064$ & $^\dagger$\cite{Bonvicini:2000qm, Nisar:2014fkc}  \\
$a_{CP}^{\mathrm{dir}}(D^+ \rightarrow \pi^0 \pi^+)$   &  $+0.029\pm 0.029$ & \cite{Mendez:2009aa} \\
$a_{CP}^{\mathrm{dir}}(D^+ \rightarrow K_S K^+)$    & $+0.0011\pm 0.0017$ &  $^\dagger$\cite{Ko:2012uh, Lees:2012jv, Mendez:2009aa, Link:2001zj,Aaij:2014qec}	\\
$a_{CP}^{\mathrm{dir}}(D_s^+\rightarrow K_S \pi^+)$  & $+0.006 \pm 0.005$  & $^\dagger$\cite{Lees:2012jv,Ko:2010ng,Mendez:2009aa,Aaij:2014qec,Aaij:2013ula}   \\
$a_{CP}^{\mathrm{dir}}(D_s^+ \rightarrow K^+ \pi^0)$ & $+0.266\pm 0.228$ &  \cite{Mendez:2009aa}  \\\hline\hline
\end{tabular}
\caption{Current data on SCS charm CP asymmetries with subtracted indirect CP violation from kaon and charm mixing \cite{Grossman:2011zk, Cenci:2012ps},
see Appendix~A of Ref.~\cite{Hiller:2012xm}. We use the notation
$\Sigma a_{CP}^{\mathrm{dir}}\equiv a_{CP}^{\mathrm{dir}}(D^0\rightarrow K^+K^-) + a_{CP}^{\mathrm{dir}}(D^0\rightarrow \pi^+\pi^-)$.
No correlations between CP asymmetries are taken into account in the fits.
$^\dagger$Our average. Table adapted from Ref.~\cite{Hiller:2014prep}.
\label{tab:cpasyms} }
\end{center}
\end{table}

We combine the sum rules Eqs.~(\ref{eq:sumrule1}) and (\ref{eq:sumrule2})
with the branching ratio fit presented in Ref.~\cite{Muller:2015lua}:
{for each point in the parameter space complying with all measured
  branching fractions (and the strong phase $\delta_{K\pi}$), we
  determine $\mathcal{S}(d)$ for the decays entering the sum rules. In the same
  step {Eqs.~(\ref{eq:defineS}),(\ref{eq:sumrule1}), and
    (\ref{eq:sumrule2})} are used to predict one CP asymmetry in terms
  of the other two.  In our fit,} we demand $\vert
{(C{+}\delta_A)}/T^{\mathrm{fac}}\vert,
\vert(E{+}\delta_A)/T^{\mathrm{fac}}\vert\leq1.3$ to {enforce} {proper}
$1/N_c$ counting.  ($T^{\mathrm{fac}}$ and $A^{\mathrm{fac}}$ are the
factorized tree and annihilation amplitudes and {$\delta_A\equiv A(D_s^+
  \rightarrow K^0 \pi^+) - A^{\mathrm{fac}}(D_s^+ \rightarrow K^0
  \pi^+)={\cal O}(1/N_c^2) $. The very conservative choice $\leq 1.3$
  accounts for a large Wilson coefficient in $E$, $C$ offsetting the
  suppression factor of $1/N_c\sim 0.3$ \cite{Muller:2015lua}.}) Apart
from the fit {with} current data {(see Table~X of
  Ref.~\cite{Muller:2015lua}),} we also consider a hypothetical future
scenario {with improved} branching ratios {by scaling their errors with}
a factor $1/\sqrt{50}$.  To illustrate the impact of the $1/N_c$
counting for the SM predictions, we perform an additional fit {without
  $1/N_c$ input.}  {This plain SU(3)$_F$ fit relies on} the topological
parametrization of Table~III in Ref.~\cite{Muller:2015lua} with the
SU(3)$_F$ counting described in Sec.~IIIB of Ref.~\cite{Muller:2015lua}.
The redundancy of the four SU(3)$_F$-limit topologies
\cite{Muller:2015lua} {is removed by absorbing $A$} into $T$, $C$, and
$E$. We {further} demand $\vert A_i^{(1)}/T\vert\leq 50\%$ {to respect
  the SU(3)$_F$ counting.}

The {experimental values} of {the} CP asymmetries included in the
fit are summarized in Table~\ref{tab:cpasyms}.  Our global fit results
are shown in Fig.~\ref{fig:sumrule}.  {The $\chi^2$ of the global minima
  range from 0.0 to 2.0 in the considered scenarios, i.e.,\ with or
  without $1/N_c$ counting and with current or future data, indicating
  an excellent fit.}  The sum rules have nontrivial implications for
direct CP asymmetries, especially when combined with the input from
$1/N_c$ counting.  With current data, we see the largest impact of the
sum rules in Fig.~\ref{plot-2d-02-26}: roughly 47\% of the 95\% C.L.
region allowed by the measurements of the CP asymmetries is excluded by
our global fit result.

Our results show that future improved measurements of branching ratios
{will} play a key role {to sharpen our} predictions: {drastic examples
  are} the prediction of $a_{CP}^{\mathrm{dir}}(D_s^+\rightarrow
K^+\pi^0)$ in Figs.~\ref{rangeplot-0-23} and \ref{rangeplot-2-23} and
the correlation of $a_{CP}^{\mathrm{dir}}(D^0\rightarrow \pi^0\pi^0)$
and $a_{CP}^{\mathrm{dir}}(D^0\rightarrow \pi^+\pi^-)$ in
Fig.~\ref{plot-2d-02-26}, where one of the two {overlapping} ellipses
vanishes in our future-data scenario. In Fig.~\ref{plot-2d-02-24}, a
smaller {portion of the experimentally allowed} region is excluded
than {in} Fig.~\ref{plot-2d-02-26}, because
$a_{CP}^{\mathrm{dir}}(D^0\to\pi^0\pi^0)$ is measured less precisely
than $\Delta a_{CP}^{\mathrm{dir}}$ (see Table~\ref{tab:cpasyms}),
{rendering the sum rule less powerful in Fig.~\ref{plot-2d-02-24}}.
{In general, the correlation of two CP asymmetries can be better
  predicted once improved data for the} third one appearing in
\eqsand{eq:sumrule1}{eq:sumrule2} {becomes} available.  {In the case of the
  fit {without $1/N_c$ counting} the {smaller errors of the} branching
  ratios in our {future scenario} do not {improve} the predictions for
  CP asymmetries.  Note that, with {current} data, the {predicted ranges}
  {barely} depend on the {additional} input from $1/N_c$ counting.}

{In principle, one can obtain the quantities {$P$} and $P+PA$, which we
  eliminate through our sum rules, from the individual CP
  asymmetries. If future data challenge our sum rules at a level which
  cannot be explained with SU(3)$_F$ breaking in {$P$} and $P+PA$, this
  will point to new physics which couples differently to $s$ and $d$
  quarks. (For an SU(3)$_F$ analysis of such models see,
  e.g.,~Ref.~\cite{Hiller:2012xm}.) However, SU(3)$_F$-symmetric new
  physics in {$P$} and $P+PA$ vanishes from the sum rules.  {(A
    similar situation can be found in the isospin sum rules of
    Ref.~\cite{Grossman:2012eb} which are insensitive to new physics in
    $\Delta I=1/2$ amplitudes.)}

\section{Conclusions}

To find {reliable} SM predictions for charm CP asymmetries, {we derive
two sum rules which treat
{$T$, $A$, $C$, $E$, $P_{\mathrm{break}}$} correctly to linear order in 
SU(3)$_F$ breaking and eliminate the penguin topologies {$P$} and $P+PA$ 
to leading order in SU(3)$_F$. Thus, we treat large tree-level parameters 
at subleading order to increase the precision in the extraction of 
loop-induced quantities sensitive to new physics. 
Unlike previously known  SU(3)$_F$-limit 
sum rules, our new sum rules correlate 
three CP asymmetries each. }
The interplay of the two sum rules probes both
the quality of SU(3)$_F$ for penguin topologies and new physics.  Future
branching ratio measurements play a key role in order to reduce the
uncertainties of the consequent SM predictions for charm CP asymmetries.

\begin{figure*}
\begin{center}
\subfigure[\, 
    {Fit including
    $1/N_c$ counting in the topological amplitudes with current (future)
    branching ratio data} in blue (green).
  {The black (magenta) line delimits the 95\% C.L. region found
  in a fit without using $1/N_c$ counting in the topological amplitudes
  for current (future) branching ratio data.} 
  Note that the black and magenta curves lie on
    top of each other.
  \label{plot-2d-02-26}]{
	\includegraphics[width=0.45\textwidth]{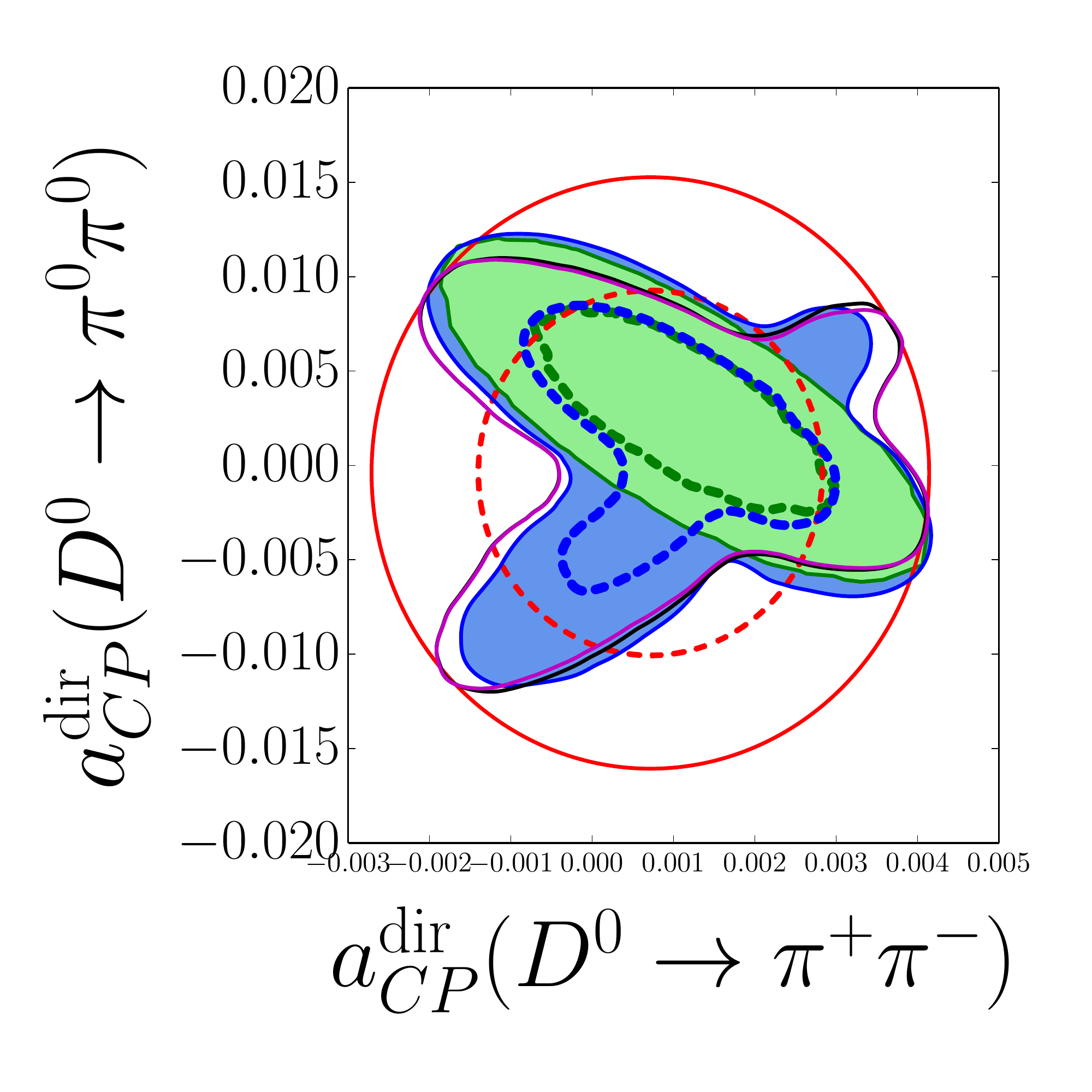} }
\hfill \subfigure[\, {Color coding as in Fig.~\ref{plot-2d-02-26}.}
The dash-dotted line denotes the
  SU(3)$_F$-limit sum rule of \eq{eq:su3limit1}.\label{plot-2d-02-24}]{
  \includegraphics[width=0.45\textwidth]{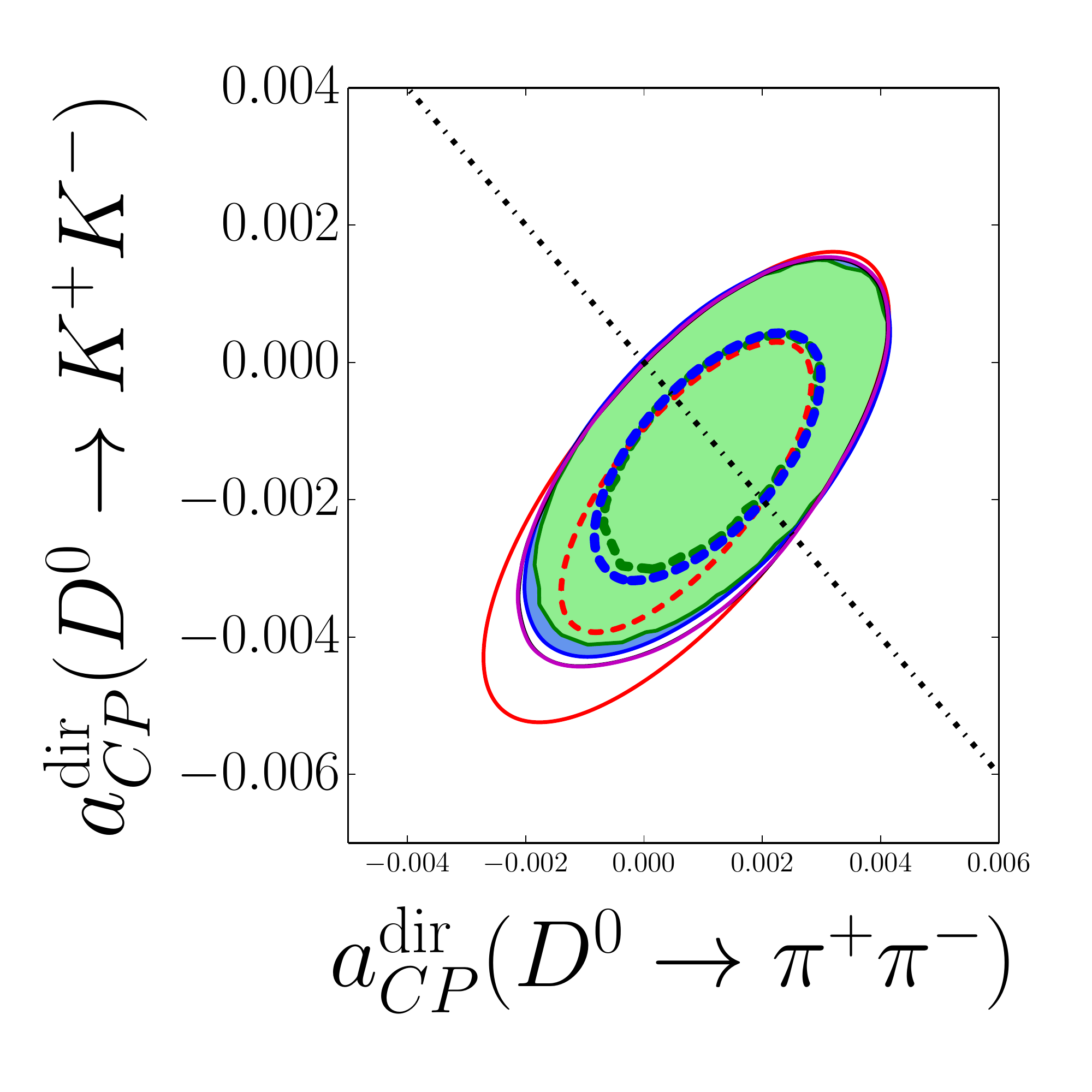} } \hfill
\begin{minipage}{0.45\textwidth}
\subfigure[\, Prediction from current branching ratio data. 
The
  dashed, solid, and dash-dotted lines correspond to 
the $1\sigma$, $2\sigma$, and $3\sigma$ intervals, respectively.
{The blue bars show our fit results for} current branching ratio data. 
{The corresponding result without $1/N_c$ counting (i.e.\ only using 
SU(3)$_F$) is shown in black.}  
\label{rangeplot-0-23}]{
	\includegraphics[width=\textwidth]{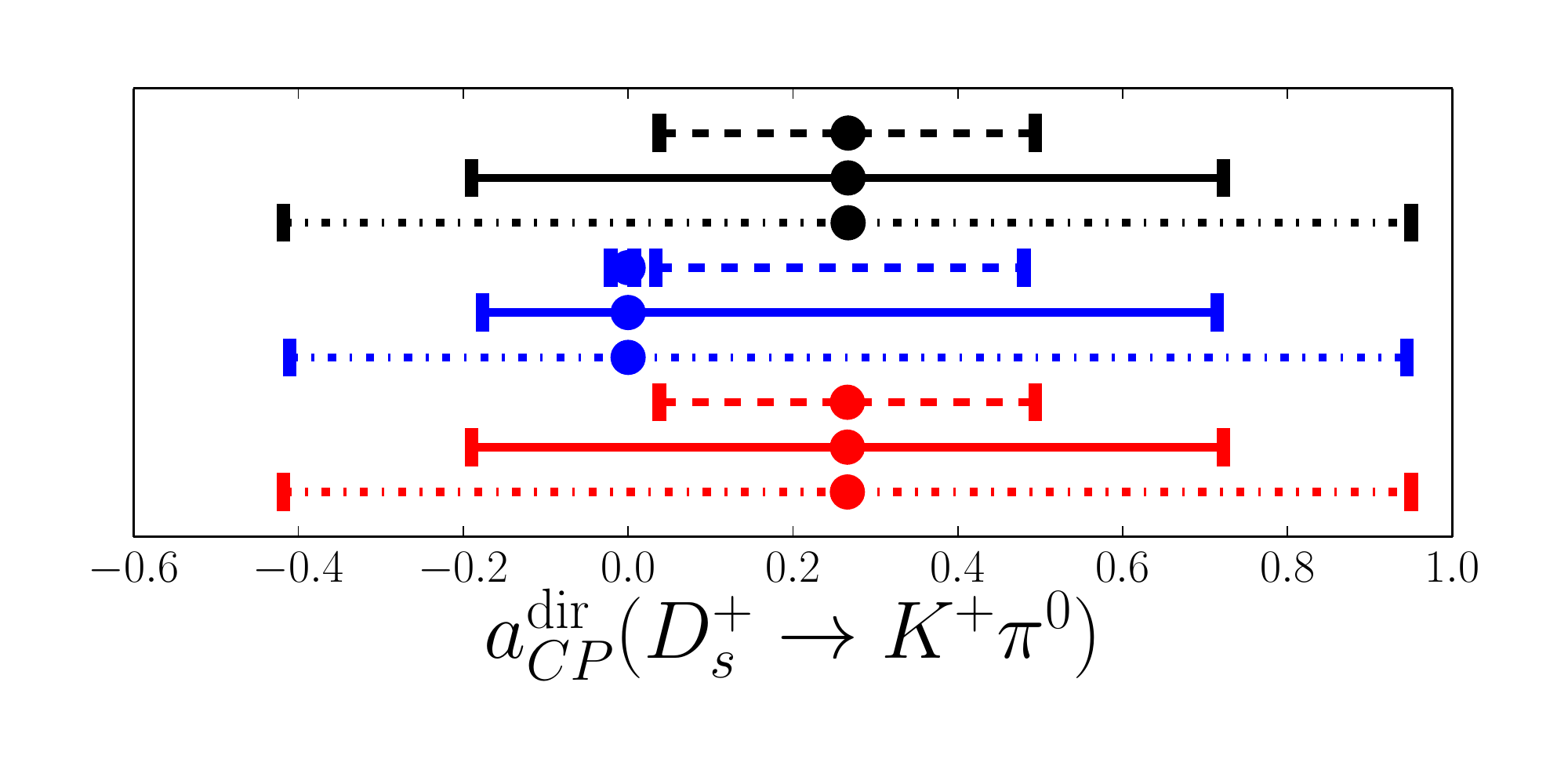} }
\subfigure[\, Same as in Fig.~(c), but for our future scenario with
  smaller errors of the branching ratios. {The green (magenta) bars 
 correspond to the analysis with (without)
    $1/N_c$ counting.}\label{rangeplot-2-23}]{
  \includegraphics[width=\textwidth]{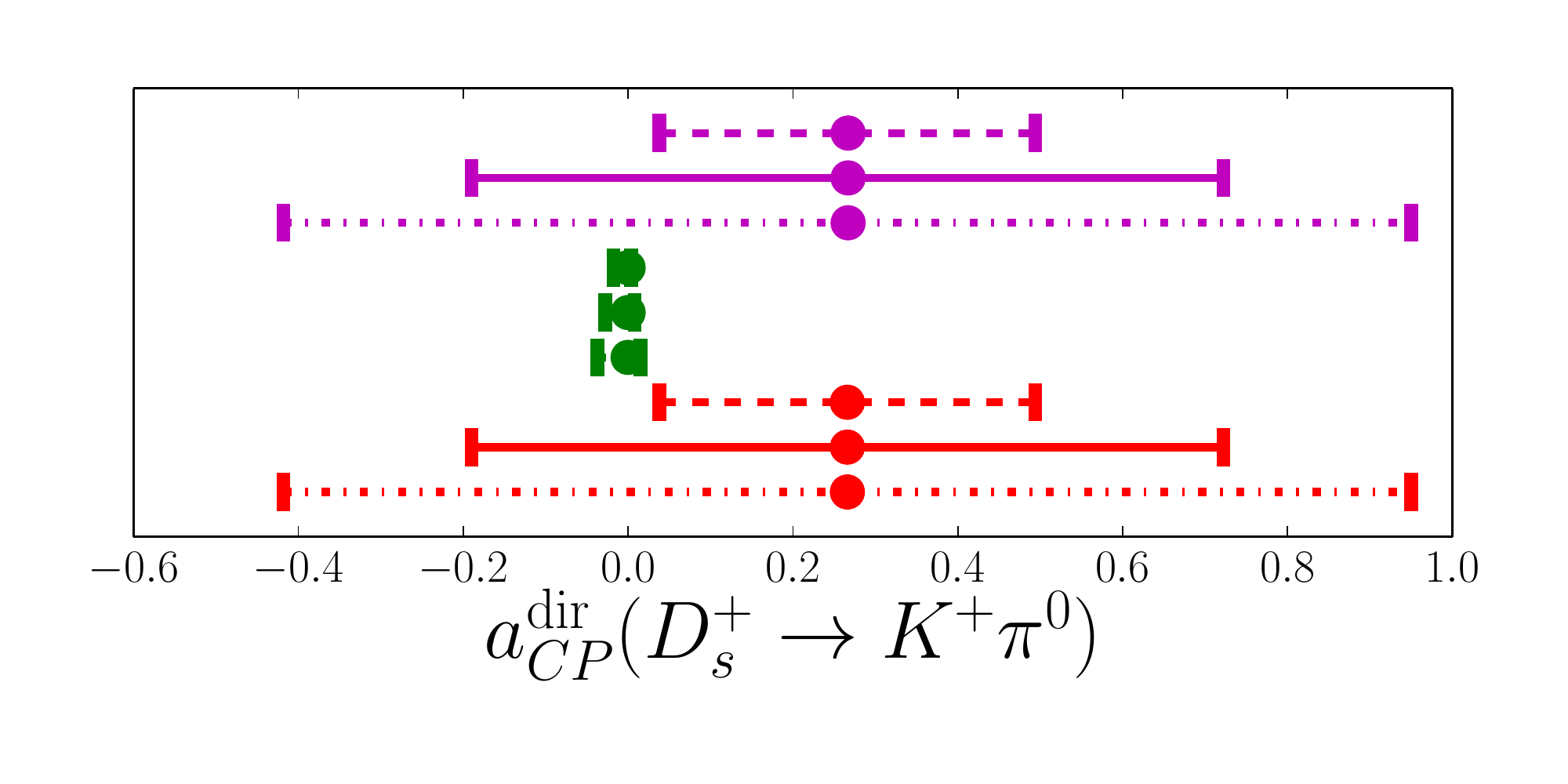} }
\end{minipage}
\begin{minipage}{0.45\textwidth}
\subfigure[\, Future scenario assuming
  $a_{CP}^{\mathrm{dir}}(D_s^+\rightarrow K^+\pi^0)=-1\%$ {and branching
    ratios with} current (future) {errors} {including $1/N_c$ counting}
  in blue (green).  The {corresponding allowed area found in a fit
    without $1/N_c$ counting is delimited by the} black (magenta) line. Note
  that the black, magenta, and red curves lie on top of each other. The
  dash-dotted line denotes the SU(3)$_F$-limit sum rule of
  \eq{eq:su3limit2}.  \label{plot-2d-47-25}]{
    \includegraphics[width=\textwidth]{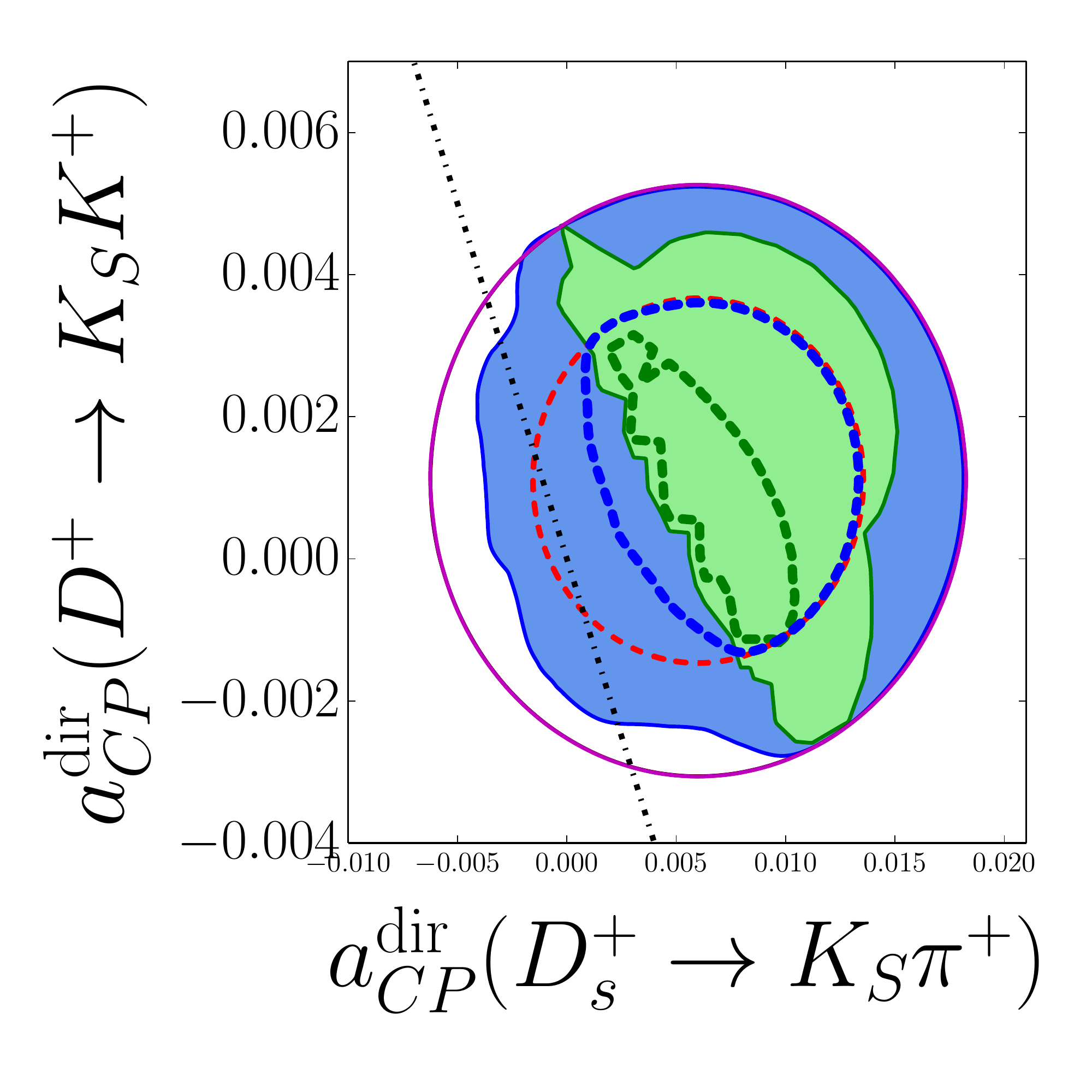} }
\end{minipage}
\end{center}
\caption{SM predictions for CP asymmetries obtained from {our}
  global fit. 
  In Figures (a),(b),(e) the dashed (solid) {red} lines {delimit
    the experimental} 68\% (95\%)~C.L. regions. {The other dashed (solid) lines are the 68\% (95\%)~C.L. regions
    of the respective fit scenarios explained in the captions of the subfigures.}
  The generic error of order $\sim 30\%$ from SU(3)$_F$ breaking in {$P$}
  and {$P+PA$} is not shown.  The experimental error ellipses are obtained
  by scaling the errors quoted in Table~\ref{tab:cpasyms} by factors of
  $\sqrt{2.28}$ and $\sqrt{5.99}$ in order to obtain the 
  two-dimensional $68\%$ and $95\%$ C.L. regions {from the
    corresponding one-dimensional ranges}. The experimental error
  ellipse shown in Fig.~\ref{plot-2d-02-24} is calculated from the
  corresponding one for $\Delta a_{CP}^{\mathrm{dir}}$ and $\Sigma
  a_{CP}^{\mathrm{dir}}$.  
\label{fig:sumrule}
}
\end{figure*}

\section{Acknowledgments}

\begin{acknowledgments}
We thank {Michael Gronau, Yuval Grossman}, Alexander Kagan, Luca
Silvestrini, Gudrun Hiller, Martin Jung, {Dean Robinson} and Jure
Zupan for useful discussions {and Michael Gronau for a critical
  proofreading of our Letter.  UN acknowledges the kind hospitality of
  the \emph{Munich Institute for Astro- and Particle Physics.}}  The
fits are performed using the \texttt{python} version of the software
package \texttt{myFitter}~\cite{Wiebusch:2012en}.  StS thanks Martin
Wiebusch for \texttt{myFitter} support and the provision of
\texttt{myFitter-python-0.2-beta}.  Parts of the computations were
performed on the {NEMO cluster of Baden-W\"urttemberg (framework
  program bwHPC).}  The presented work is supported by BMBF under
contract no.~05H12VKF.
\end{acknowledgments}


\bibliography{mns-14-cpv.bib}

\begin{thebibliography}{64}%
\makeatletter
\providecommand \@ifxundefined [1]{%
 \@ifx{#1\undefined}
}%
\providecommand \@ifnum [1]{%
 \ifnum #1\expandafter \@firstoftwo
 \else \expandafter \@secondoftwo
 \fi
}%
\providecommand \@ifx [1]{%
 \ifx #1\expandafter \@firstoftwo
 \else \expandafter \@secondoftwo
 \fi
}%
\providecommand \natexlab [1]{#1}%
\providecommand \enquote  [1]{``#1''}%
\providecommand \bibnamefont  [1]{#1}%
\providecommand \bibfnamefont [1]{#1}%
\providecommand \citenamefont [1]{#1}%
\providecommand \href@noop [0]{\@secondoftwo}%
\providecommand \href [0]{\begingroup \@sanitize@url \@href}%
\providecommand \@href[1]{\@@startlink{#1}\@@href}%
\providecommand \@@href[1]{\endgroup#1\@@endlink}%
\providecommand \@sanitize@url [0]{\catcode `\\12\catcode `\$12\catcode
  `\&12\catcode `\#12\catcode `\^12\catcode `\_12\catcode `\%12\relax}%
\providecommand \@@startlink[1]{}%
\providecommand \@@endlink[0]{}%
\providecommand \url  [0]{\begingroup\@sanitize@url \@url }%
\providecommand \@url [1]{\endgroup\@href {#1}{\urlprefix }}%
\providecommand \urlprefix  [0]{URL }%
\providecommand \Eprint [0]{\href }%
\providecommand \doibase [0]{http://dx.doi.org/}%
\providecommand \selectlanguage [0]{\@gobble}%
\providecommand \bibinfo  [0]{\@secondoftwo}%
\providecommand \bibfield  [0]{\@secondoftwo}%
\providecommand \translation [1]{[#1]}%
\providecommand \BibitemOpen [0]{}%
\providecommand \bibitemStop [0]{}%
\providecommand \bibitemNoStop [0]{.\EOS\space}%
\providecommand \EOS [0]{\spacefactor3000\relax}%
\providecommand \BibitemShut  [1]{\csname bibitem#1\endcsname}%
\let\auto@bib@innerbib\@empty
\bibitem [{\citenamefont {Aaij}\ \emph {et~al.}(2012)\citenamefont {Aaij} \emph
  {et~al.}}]{Aaij:2011in}%
  \BibitemOpen
  \bibfield  {author} {\bibinfo {author} {\bibfnamefont {R.}~\bibnamefont
  {Aaij}} \emph {et~al.} (\bibinfo {collaboration} {LHCb Collaboration}),\
  }\href {\doibase 10.1103/PhysRevLett.108.129903,
  10.1103/PhysRevLett.108.111602} {\bibfield  {journal} {\bibinfo  {journal}
  {Phys.Rev.Lett.}\ }\textbf {\bibinfo {volume} {108}},\ \bibinfo {pages}
  {111602} (\bibinfo {year} {2012})},\ \Eprint {http://arxiv.org/abs/1112.0938}
  {arXiv:1112.0938 [hep-ex]} \BibitemShut {NoStop}%
\bibitem [{\citenamefont {Aaltonen}\ \emph
  {et~al.}(2012{\natexlab{a}})\citenamefont {Aaltonen} \emph
  {et~al.}}]{Collaboration:2012qw}%
  \BibitemOpen
  \bibfield  {author} {\bibinfo {author} {\bibfnamefont {T.}~\bibnamefont
  {Aaltonen}} \emph {et~al.} (\bibinfo {collaboration} {CDF Collaboration}),\
  }\href {\doibase 10.1103/PhysRevLett.109.111801} {\bibfield  {journal}
  {\bibinfo  {journal} {Phys.Rev.Lett.}\ }\textbf {\bibinfo {volume} {109}},\
  \bibinfo {pages} {111801} (\bibinfo {year} {2012}{\natexlab{a}})},\ \Eprint
  {http://arxiv.org/abs/1207.2158} {arXiv:1207.2158 [hep-ex]} \BibitemShut
  {NoStop}%
\bibitem [{\citenamefont {Aaij}\ \emph
  {et~al.}(2014{\natexlab{a}})\citenamefont {Aaij} \emph
  {et~al.}}]{Aaij:2014gsa}%
  \BibitemOpen
  \bibfield  {author} {\bibinfo {author} {\bibfnamefont {R.}~\bibnamefont
  {Aaij}} \emph {et~al.} (\bibinfo {collaboration} {LHCb collaboration}),\
  }\href {\doibase 10.1007/JHEP07(2014)041} {\bibfield  {journal} {\bibinfo
  {journal} {JHEP}\ }\textbf {\bibinfo {volume} {1407}},\ \bibinfo {pages}
  {041} (\bibinfo {year} {2014}{\natexlab{a}})},\ \Eprint
  {http://arxiv.org/abs/1405.2797} {arXiv:1405.2797 [hep-ex]} \BibitemShut
  {NoStop}%
\bibitem [{\citenamefont {Nisar}\ \emph {et~al.}(2014)\citenamefont {Nisar}
  \emph {et~al.}}]{Nisar:2014fkc}%
  \BibitemOpen
  \bibfield  {author} {\bibinfo {author} {\bibfnamefont {N.}~\bibnamefont
  {Nisar}} \emph {et~al.} (\bibinfo {collaboration} {Belle Collaboration}),\
  }\href {\doibase 10.1103/PhysRevLett.112.211601} {\bibfield  {journal}
  {\bibinfo  {journal} {Phys.Rev.Lett.}\ }\textbf {\bibinfo {volume} {112}},\
  \bibinfo {pages} {211601} (\bibinfo {year} {2014})},\ \Eprint
  {http://arxiv.org/abs/1404.1266} {arXiv:1404.1266 [hep-ex]} \BibitemShut
  {NoStop}%
\bibitem [{\citenamefont {Bigi}\ \emph {et~al.}(2011)\citenamefont {Bigi},
  \citenamefont {Paul},\ and\ \citenamefont {Recksiegel}}]{Bigi:2011re}%
  \BibitemOpen
  \bibfield  {author} {\bibinfo {author} {\bibfnamefont {I.~I.}\ \bibnamefont
  {Bigi}}, \bibinfo {author} {\bibfnamefont {A.}~\bibnamefont {Paul}}, \ and\
  \bibinfo {author} {\bibfnamefont {S.}~\bibnamefont {Recksiegel}},\ }\href
  {\doibase 10.1007/JHEP06(2011)089} {\bibfield  {journal} {\bibinfo  {journal}
  {JHEP}\ }\textbf {\bibinfo {volume} {1106}},\ \bibinfo {pages} {089}
  (\bibinfo {year} {2011})},\ \Eprint {http://arxiv.org/abs/1103.5785}
  {arXiv:1103.5785 [hep-ph]} \BibitemShut {NoStop}%
\bibitem [{\citenamefont {Buccella}\ \emph {et~al.}(1995)\citenamefont
  {Buccella}, \citenamefont {Lusignoli}, \citenamefont {Miele}, \citenamefont
  {Pugliese},\ and\ \citenamefont {Santorelli}}]{Buccella:1994nf}%
  \BibitemOpen
  \bibfield  {author} {\bibinfo {author} {\bibfnamefont {F.}~\bibnamefont
  {Buccella}}, \bibinfo {author} {\bibfnamefont {M.}~\bibnamefont {Lusignoli}},
  \bibinfo {author} {\bibfnamefont {G.}~\bibnamefont {Miele}}, \bibinfo
  {author} {\bibfnamefont {A.}~\bibnamefont {Pugliese}}, \ and\ \bibinfo
  {author} {\bibfnamefont {P.}~\bibnamefont {Santorelli}},\ }\href {\doibase
  10.1103/PhysRevD.51.3478} {\bibfield  {journal} {\bibinfo  {journal}
  {Phys.Rev.}\ }\textbf {\bibinfo {volume} {D51}},\ \bibinfo {pages} {3478}
  (\bibinfo {year} {1995})},\ \Eprint {http://arxiv.org/abs/hep-ph/9411286}
  {arXiv:hep-ph/9411286 [hep-ph]} \BibitemShut {NoStop}%
\bibitem [{\citenamefont {Grossman}\ \emph {et~al.}(2007)\citenamefont
  {Grossman}, \citenamefont {Kagan},\ and\ \citenamefont
  {Nir}}]{Grossman:2006jg}%
  \BibitemOpen
  \bibfield  {author} {\bibinfo {author} {\bibfnamefont {Y.}~\bibnamefont
  {Grossman}}, \bibinfo {author} {\bibfnamefont {A.~L.}\ \bibnamefont {Kagan}},
  \ and\ \bibinfo {author} {\bibfnamefont {Y.}~\bibnamefont {Nir}},\ }\href
  {\doibase 10.1103/PhysRevD.75.036008} {\bibfield  {journal} {\bibinfo
  {journal} {Phys.Rev.}\ }\textbf {\bibinfo {volume} {D75}},\ \bibinfo {pages}
  {036008} (\bibinfo {year} {2007})},\ \Eprint
  {http://arxiv.org/abs/hep-ph/0609178} {arXiv:hep-ph/0609178 [hep-ph]}
  \BibitemShut {NoStop}%
\bibitem [{\citenamefont {Artuso}\ \emph {et~al.}(2008)\citenamefont {Artuso},
  \citenamefont {Meadows},\ and\ \citenamefont {Petrov}}]{Artuso:2008vf}%
  \BibitemOpen
  \bibfield  {author} {\bibinfo {author} {\bibfnamefont {M.}~\bibnamefont
  {Artuso}}, \bibinfo {author} {\bibfnamefont {B.}~\bibnamefont {Meadows}}, \
  and\ \bibinfo {author} {\bibfnamefont {A.~A.}\ \bibnamefont {Petrov}},\
  }\href {\doibase 10.1146/annurev.nucl.58.110707.171131} {\bibfield  {journal}
  {\bibinfo  {journal} {Ann.Rev.Nucl.Part.Sci.}\ }\textbf {\bibinfo {volume}
  {58}},\ \bibinfo {pages} {249} (\bibinfo {year} {2008})},\ \Eprint
  {http://arxiv.org/abs/0802.2934} {arXiv:0802.2934 [hep-ph]} \BibitemShut
  {NoStop}%
\bibitem [{\citenamefont {Petrov}(2009)}]{Petrov:2010gy}%
  \BibitemOpen
  \bibfield  {author} {\bibinfo {author} {\bibfnamefont {A.~A.}\ \bibnamefont
  {Petrov}},\ }\href@noop {} {\bibfield  {journal} {\bibinfo  {journal} {PoS}\
  }\textbf {\bibinfo {volume} {BEAUTY2009}},\ \bibinfo {pages} {024} (\bibinfo
  {year} {2009})},\ \Eprint {http://arxiv.org/abs/1003.0906} {arXiv:1003.0906
  [hep-ph]} \BibitemShut {NoStop}%
\bibitem [{\citenamefont {Li}\ \emph {et~al.}(2012)\citenamefont {Li},
  \citenamefont {Lu},\ and\ \citenamefont {Yu}}]{Li:2012cfa}%
  \BibitemOpen
  \bibfield  {author} {\bibinfo {author} {\bibfnamefont {H.-n.}\ \bibnamefont
  {Li}}, \bibinfo {author} {\bibfnamefont {C.-D.}\ \bibnamefont {Lu}}, \ and\
  \bibinfo {author} {\bibfnamefont {F.-S.}\ \bibnamefont {Yu}},\ }\href
  {\doibase 10.1103/PhysRevD.86.036012} {\bibfield  {journal} {\bibinfo
  {journal} {Phys.Rev.}\ }\textbf {\bibinfo {volume} {D86}},\ \bibinfo {pages}
  {036012} (\bibinfo {year} {2012})},\ \Eprint {http://arxiv.org/abs/1203.3120}
  {arXiv:1203.3120 [hep-ph]} \BibitemShut {NoStop}%
\bibitem [{\citenamefont {Cheng}\ and\ \citenamefont
  {Chiang}(2012)}]{Cheng:2012wr}%
  \BibitemOpen
  \bibfield  {author} {\bibinfo {author} {\bibfnamefont {H.-Y.}\ \bibnamefont
  {Cheng}}\ and\ \bibinfo {author} {\bibfnamefont {C.-W.}\ \bibnamefont
  {Chiang}},\ }\href {\doibase 10.1103/PhysRevD.85.079903,
  10.1103/PhysRevD.85.034036} {\bibfield  {journal} {\bibinfo  {journal}
  {Phys.Rev.}\ }\textbf {\bibinfo {volume} {D85}},\ \bibinfo {pages} {034036}
  (\bibinfo {year} {2012})},\ \Eprint {http://arxiv.org/abs/1201.0785}
  {arXiv:1201.0785 [hep-ph]} \BibitemShut {NoStop}%
\bibitem [{\citenamefont {Brod}\ \emph
  {et~al.}(2012{\natexlab{a}})\citenamefont {Brod}, \citenamefont {Kagan},\
  and\ \citenamefont {Zupan}}]{Brod:2011re}%
  \BibitemOpen
  \bibfield  {author} {\bibinfo {author} {\bibfnamefont {J.}~\bibnamefont
  {Brod}}, \bibinfo {author} {\bibfnamefont {A.~L.}\ \bibnamefont {Kagan}}, \
  and\ \bibinfo {author} {\bibfnamefont {J.}~\bibnamefont {Zupan}},\ }\href
  {\doibase 10.1103/PhysRevD.86.014023} {\bibfield  {journal} {\bibinfo
  {journal} {Phys.Rev.}\ }\textbf {\bibinfo {volume} {D86}},\ \bibinfo {pages}
  {014023} (\bibinfo {year} {2012}{\natexlab{a}})},\ \Eprint
  {http://arxiv.org/abs/1111.5000} {arXiv:1111.5000 [hep-ph]} \BibitemShut
  {NoStop}%
\bibitem [{\citenamefont {Pirtskhalava}\ and\ \citenamefont
  {Uttayarat}(2012)}]{Pirtskhalava:2011va}%
  \BibitemOpen
  \bibfield  {author} {\bibinfo {author} {\bibfnamefont {D.}~\bibnamefont
  {Pirtskhalava}}\ and\ \bibinfo {author} {\bibfnamefont {P.}~\bibnamefont
  {Uttayarat}},\ }\href {\doibase 10.1016/j.physletb.2012.04.039} {\bibfield
  {journal} {\bibinfo  {journal} {Phys.Lett.}\ }\textbf {\bibinfo {volume}
  {B712}},\ \bibinfo {pages} {81} (\bibinfo {year} {2012})},\ \Eprint
  {http://arxiv.org/abs/1112.5451} {arXiv:1112.5451 [hep-ph]} \BibitemShut
  {NoStop}%
\bibitem [{\citenamefont {Brod}\ \emph
  {et~al.}(2012{\natexlab{b}})\citenamefont {Brod}, \citenamefont {Grossman},
  \citenamefont {Kagan},\ and\ \citenamefont {Zupan}}]{Brod:2012ud}%
  \BibitemOpen
  \bibfield  {author} {\bibinfo {author} {\bibfnamefont {J.}~\bibnamefont
  {Brod}}, \bibinfo {author} {\bibfnamefont {Y.}~\bibnamefont {Grossman}},
  \bibinfo {author} {\bibfnamefont {A.~L.}\ \bibnamefont {Kagan}}, \ and\
  \bibinfo {author} {\bibfnamefont {J.}~\bibnamefont {Zupan}},\ }\href
  {\doibase 10.1007/JHEP10(2012)161} {\bibfield  {journal} {\bibinfo  {journal}
  {JHEP}\ }\textbf {\bibinfo {volume} {1210}},\ \bibinfo {pages} {161}
  (\bibinfo {year} {2012}{\natexlab{b}})},\ \Eprint
  {http://arxiv.org/abs/1203.6659} {arXiv:1203.6659 [hep-ph]} \BibitemShut
  {NoStop}%
\bibitem [{\citenamefont {Feldmann}\ \emph {et~al.}(2012)\citenamefont
  {Feldmann}, \citenamefont {Nandi},\ and\ \citenamefont
  {Soni}}]{Feldmann:2012js}%
  \BibitemOpen
  \bibfield  {author} {\bibinfo {author} {\bibfnamefont {T.}~\bibnamefont
  {Feldmann}}, \bibinfo {author} {\bibfnamefont {S.}~\bibnamefont {Nandi}}, \
  and\ \bibinfo {author} {\bibfnamefont {A.}~\bibnamefont {Soni}},\ }\href
  {\doibase 10.1007/JHEP06(2012)007} {\bibfield  {journal} {\bibinfo  {journal}
  {JHEP}\ }\textbf {\bibinfo {volume} {1206}},\ \bibinfo {pages} {007}
  (\bibinfo {year} {2012})},\ \Eprint {http://arxiv.org/abs/1202.3795}
  {arXiv:1202.3795 [hep-ph]} \BibitemShut {NoStop}%
\bibitem [{\citenamefont {Hiller}\ \emph {et~al.}(2013)\citenamefont {Hiller},
  \citenamefont {Jung},\ and\ \citenamefont {Schacht}}]{Hiller:2012xm}%
  \BibitemOpen
  \bibfield  {author} {\bibinfo {author} {\bibfnamefont {G.}~\bibnamefont
  {Hiller}}, \bibinfo {author} {\bibfnamefont {M.}~\bibnamefont {Jung}}, \ and\
  \bibinfo {author} {\bibfnamefont {S.}~\bibnamefont {Schacht}},\ }\href
  {\doibase 10.1103/PhysRevD.87.014024} {\bibfield  {journal} {\bibinfo
  {journal} {Phys.Rev.}\ }\textbf {\bibinfo {volume} {D87}},\ \bibinfo {pages}
  {014024} (\bibinfo {year} {2013})},\ \Eprint {http://arxiv.org/abs/1211.3734}
  {arXiv:1211.3734 [hep-ph]} \BibitemShut {NoStop}%
\bibitem [{\citenamefont {Franco}\ \emph {et~al.}(2012)\citenamefont {Franco},
  \citenamefont {Mishima},\ and\ \citenamefont {Silvestrini}}]{Franco:2012ck}%
  \BibitemOpen
  \bibfield  {author} {\bibinfo {author} {\bibfnamefont {E.}~\bibnamefont
  {Franco}}, \bibinfo {author} {\bibfnamefont {S.}~\bibnamefont {Mishima}}, \
  and\ \bibinfo {author} {\bibfnamefont {L.}~\bibnamefont {Silvestrini}},\
  }\href {\doibase 10.1007/JHEP05(2012)140} {\bibfield  {journal} {\bibinfo
  {journal} {JHEP}\ }\textbf {\bibinfo {volume} {1205}},\ \bibinfo {pages}
  {140} (\bibinfo {year} {2012})},\ \Eprint {http://arxiv.org/abs/1203.3131}
  {arXiv:1203.3131 [hep-ph]} \BibitemShut {NoStop}%
\bibitem [{\citenamefont {Golden}\ and\ \citenamefont
  {Grinstein}(1989)}]{Golden:1989qx}%
  \BibitemOpen
  \bibfield  {author} {\bibinfo {author} {\bibfnamefont {M.}~\bibnamefont
  {Golden}}\ and\ \bibinfo {author} {\bibfnamefont {B.}~\bibnamefont
  {Grinstein}},\ }\href {\doibase 10.1016/0370-2693(89)90353-5} {\bibfield
  {journal} {\bibinfo  {journal} {Phys.Lett.}\ }\textbf {\bibinfo {volume}
  {B222}},\ \bibinfo {pages} {501} (\bibinfo {year} {1989})}\BibitemShut
  {NoStop}%
\bibitem [{\citenamefont {Bobrowski}\ \emph {et~al.}(2010)\citenamefont
  {Bobrowski}, \citenamefont {Lenz}, \citenamefont {Riedl},\ and\ \citenamefont
  {Rohrwild}}]{Bobrowski:2010xg}%
  \BibitemOpen
  \bibfield  {author} {\bibinfo {author} {\bibfnamefont {M.}~\bibnamefont
  {Bobrowski}}, \bibinfo {author} {\bibfnamefont {A.}~\bibnamefont {Lenz}},
  \bibinfo {author} {\bibfnamefont {J.}~\bibnamefont {Riedl}}, \ and\ \bibinfo
  {author} {\bibfnamefont {J.}~\bibnamefont {Rohrwild}},\ }\href {\doibase
  10.1007/JHEP03(2010)009} {\bibfield  {journal} {\bibinfo  {journal} {JHEP}\
  }\textbf {\bibinfo {volume} {1003}},\ \bibinfo {pages} {009} (\bibinfo {year}
  {2010})},\ \Eprint {http://arxiv.org/abs/1002.4794} {arXiv:1002.4794
  [hep-ph]} \BibitemShut {NoStop}%
\bibitem [{\citenamefont {Altarelli}\ \emph {et~al.}(1975)\citenamefont
  {Altarelli}, \citenamefont {Cabibbo},\ and\ \citenamefont
  {Maiani}}]{Altarelli:1974sc}%
  \BibitemOpen
  \bibfield  {author} {\bibinfo {author} {\bibfnamefont {G.}~\bibnamefont
  {Altarelli}}, \bibinfo {author} {\bibfnamefont {N.}~\bibnamefont {Cabibbo}},
  \ and\ \bibinfo {author} {\bibfnamefont {L.}~\bibnamefont {Maiani}},\ }\href
  {\doibase 10.1016/0550-3213(75)90281-3} {\bibfield  {journal} {\bibinfo
  {journal} {Nucl.Phys.}\ }\textbf {\bibinfo {volume} {B88}},\ \bibinfo {pages}
  {285} (\bibinfo {year} {1975})}\BibitemShut {NoStop}%
\bibitem [{\citenamefont {Kingsley}\ \emph {et~al.}(1975)\citenamefont
  {Kingsley}, \citenamefont {Treiman}, \citenamefont {Wilczek},\ and\
  \citenamefont {Zee}}]{Kingsley:1975fe}%
  \BibitemOpen
  \bibfield  {author} {\bibinfo {author} {\bibfnamefont {R.}~\bibnamefont
  {Kingsley}}, \bibinfo {author} {\bibfnamefont {S.}~\bibnamefont {Treiman}},
  \bibinfo {author} {\bibfnamefont {F.}~\bibnamefont {Wilczek}}, \ and\
  \bibinfo {author} {\bibfnamefont {A.}~\bibnamefont {Zee}},\ }\href {\doibase
  10.1103/PhysRevD.11.1919} {\bibfield  {journal} {\bibinfo  {journal}
  {Phys.Rev.}\ }\textbf {\bibinfo {volume} {D11}},\ \bibinfo {pages} {1919}
  (\bibinfo {year} {1975})}\BibitemShut {NoStop}%
\bibitem [{\citenamefont {Einhorn}\ and\ \citenamefont
  {Quigg}(1975)}]{Einhorn:1975fw}%
  \BibitemOpen
  \bibfield  {author} {\bibinfo {author} {\bibfnamefont {M.}~\bibnamefont
  {Einhorn}}\ and\ \bibinfo {author} {\bibfnamefont {C.}~\bibnamefont
  {Quigg}},\ }\href {\doibase 10.1103/PhysRevD.12.2015} {\bibfield  {journal}
  {\bibinfo  {journal} {Phys.Rev.}\ }\textbf {\bibinfo {volume} {D12}},\
  \bibinfo {pages} {2015} (\bibinfo {year} {1975})}\BibitemShut {NoStop}%
\bibitem [{\citenamefont {Voloshin}\ \emph {et~al.}(1975)\citenamefont
  {Voloshin}, \citenamefont {Zakharov},\ and\ \citenamefont
  {Okun}}]{Voloshin:1975yx}%
  \BibitemOpen
  \bibfield  {author} {\bibinfo {author} {\bibfnamefont {M.}~\bibnamefont
  {Voloshin}}, \bibinfo {author} {\bibfnamefont {V.}~\bibnamefont {Zakharov}},
  \ and\ \bibinfo {author} {\bibfnamefont {L.}~\bibnamefont {Okun}},\
  }\href@noop {} {\bibfield  {journal} {\bibinfo  {journal} {JETP Lett.}\
  }\textbf {\bibinfo {volume} {21}},\ \bibinfo {pages} {183} (\bibinfo {year}
  {1975})}\BibitemShut {NoStop}%
\bibitem [{\citenamefont {Quigg}(1980)}]{Quigg:1979ic}%
  \BibitemOpen
  \bibfield  {author} {\bibinfo {author} {\bibfnamefont {C.}~\bibnamefont
  {Quigg}},\ }\href {\doibase 10.1007/BF01477308} {\bibfield  {journal}
  {\bibinfo  {journal} {Z.Phys.}\ }\textbf {\bibinfo {volume} {C4}},\ \bibinfo
  {pages} {55} (\bibinfo {year} {1980})}\BibitemShut {NoStop}%
\bibitem [{\citenamefont {{L.-L. Chau Wang}}(1980)}]{Wang:1980ac}%
  \BibitemOpen
  \bibfield  {author} {\bibinfo {author} {\bibnamefont {{L.-L. Chau Wang}}},\
  }\href@noop {} {\enquote {\bibinfo {title} {{Flavor Mixing and Charm
  Decay}},}\ } (\bibinfo {year} {1980}),\ \bibinfo {note} {{BNL-27615,
  C80-01-05-20, Talk at the Conference on Theoretical Particle Physics,
  5-14~January~1980, Guangzhou (Canton), China, p.~1218}}\BibitemShut {NoStop}%
\bibitem [{\citenamefont {Chau}(1983)}]{Chau:1982da}%
  \BibitemOpen
  \bibfield  {author} {\bibinfo {author} {\bibfnamefont {L.-L.}\ \bibnamefont
  {Chau}},\ }\href {\doibase 10.1016/0370-1573(83)90043-1} {\bibfield
  {journal} {\bibinfo  {journal} {Phys.Rept.}\ }\textbf {\bibinfo {volume}
  {95}},\ \bibinfo {pages} {1} (\bibinfo {year} {1983})}\BibitemShut {NoStop}%
\bibitem [{\citenamefont {Savage}(1991)}]{Savage:1991wu}%
  \BibitemOpen
  \bibfield  {author} {\bibinfo {author} {\bibfnamefont {M.~J.}\ \bibnamefont
  {Savage}},\ }\href {\doibase 10.1016/0370-2693(91)91917-K} {\bibfield
  {journal} {\bibinfo  {journal} {Phys.Lett.}\ }\textbf {\bibinfo {volume}
  {B257}},\ \bibinfo {pages} {414} (\bibinfo {year} {1991})}\BibitemShut
  {NoStop}%
\bibitem [{\citenamefont {Chau}\ and\ \citenamefont
  {Cheng}(1992)}]{Chau:1991gx}%
  \BibitemOpen
  \bibfield  {author} {\bibinfo {author} {\bibfnamefont {L.-L.}\ \bibnamefont
  {Chau}}\ and\ \bibinfo {author} {\bibfnamefont {H.-Y.}\ \bibnamefont
  {Cheng}},\ }\href {\doibase 10.1016/0370-2693(92)90067-E} {\bibfield
  {journal} {\bibinfo  {journal} {Phys.Lett.}\ }\textbf {\bibinfo {volume}
  {B280}},\ \bibinfo {pages} {281} (\bibinfo {year} {1992})}\BibitemShut
  {NoStop}%
\bibitem [{\citenamefont {Isidori}\ \emph {et~al.}(2012)\citenamefont
  {Isidori}, \citenamefont {Kamenik}, \citenamefont {Ligeti},\ and\
  \citenamefont {Perez}}]{Isidori:2011qw}%
  \BibitemOpen
  \bibfield  {author} {\bibinfo {author} {\bibfnamefont {G.}~\bibnamefont
  {Isidori}}, \bibinfo {author} {\bibfnamefont {J.~F.}\ \bibnamefont
  {Kamenik}}, \bibinfo {author} {\bibfnamefont {Z.}~\bibnamefont {Ligeti}}, \
  and\ \bibinfo {author} {\bibfnamefont {G.}~\bibnamefont {Perez}},\ }\href
  {\doibase 10.1016/j.physletb.2012.03.046} {\bibfield  {journal} {\bibinfo
  {journal} {Phys.Lett.}\ }\textbf {\bibinfo {volume} {B711}},\ \bibinfo
  {pages} {46} (\bibinfo {year} {2012})},\ \Eprint
  {http://arxiv.org/abs/1111.4987} {arXiv:1111.4987 [hep-ph]} \BibitemShut
  {NoStop}%
\bibitem [{\citenamefont {Bhattacharya}\ \emph
  {et~al.}(2012{\natexlab{a}})\citenamefont {Bhattacharya}, \citenamefont
  {Gronau},\ and\ \citenamefont {Rosner}}]{Bhattacharya:2012ah}%
  \BibitemOpen
  \bibfield  {author} {\bibinfo {author} {\bibfnamefont {B.}~\bibnamefont
  {Bhattacharya}}, \bibinfo {author} {\bibfnamefont {M.}~\bibnamefont
  {Gronau}}, \ and\ \bibinfo {author} {\bibfnamefont {J.~L.}\ \bibnamefont
  {Rosner}},\ }\href {\doibase 10.1103/PhysRevD.85.079901,
  10.1103/PhysRevD.85.054014} {\bibfield  {journal} {\bibinfo  {journal}
  {Phys.Rev.}\ }\textbf {\bibinfo {volume} {D85}},\ \bibinfo {pages} {054014}
  (\bibinfo {year} {2012}{\natexlab{a}})},\ \Eprint
  {http://arxiv.org/abs/1201.2351} {arXiv:1201.2351 [hep-ph]} \BibitemShut
  {NoStop}%
\bibitem [{\citenamefont {Altmannshofer}\ \emph {et~al.}(2012)\citenamefont
  {Altmannshofer}, \citenamefont {Primulando}, \citenamefont {Yu},\ and\
  \citenamefont {Yu}}]{Altmannshofer:2012ur}%
  \BibitemOpen
  \bibfield  {author} {\bibinfo {author} {\bibfnamefont {W.}~\bibnamefont
  {Altmannshofer}}, \bibinfo {author} {\bibfnamefont {R.}~\bibnamefont
  {Primulando}}, \bibinfo {author} {\bibfnamefont {C.-T.}\ \bibnamefont {Yu}},
  \ and\ \bibinfo {author} {\bibfnamefont {F.}~\bibnamefont {Yu}},\ }\href
  {\doibase 10.1007/JHEP04(2012)049} {\bibfield  {journal} {\bibinfo  {journal}
  {JHEP}\ }\textbf {\bibinfo {volume} {1204}},\ \bibinfo {pages} {049}
  (\bibinfo {year} {2012})},\ \Eprint {http://arxiv.org/abs/1202.2866}
  {arXiv:1202.2866 [hep-ph]} \BibitemShut {NoStop}%
\bibitem [{\citenamefont {Bhattacharya}\ \emph
  {et~al.}(2012{\natexlab{b}})\citenamefont {Bhattacharya}, \citenamefont
  {Gronau},\ and\ \citenamefont {Rosner}}]{Bhattacharya:2012pc}%
  \BibitemOpen
  \bibfield  {author} {\bibinfo {author} {\bibfnamefont {B.}~\bibnamefont
  {Bhattacharya}}, \bibinfo {author} {\bibfnamefont {M.}~\bibnamefont
  {Gronau}}, \ and\ \bibinfo {author} {\bibfnamefont {J.~L.}\ \bibnamefont
  {Rosner}},\ }\href@noop {} {\  (\bibinfo {year} {2012}{\natexlab{b}})},\
  \Eprint {http://arxiv.org/abs/1207.6390} {arXiv:1207.6390 [hep-ph]}
  \BibitemShut {NoStop}%
\bibitem [{\citenamefont {Bhattacharya}\ \emph
  {et~al.}(2012{\natexlab{c}})\citenamefont {Bhattacharya}, \citenamefont
  {Gronau},\ and\ \citenamefont {Rosner}}]{Bhattacharya:2012kq}%
  \BibitemOpen
  \bibfield  {author} {\bibinfo {author} {\bibfnamefont {B.}~\bibnamefont
  {Bhattacharya}}, \bibinfo {author} {\bibfnamefont {M.}~\bibnamefont
  {Gronau}}, \ and\ \bibinfo {author} {\bibfnamefont {J.~L.}\ \bibnamefont
  {Rosner}},\ }\href@noop {} {\  (\bibinfo {year} {2012}{\natexlab{c}})},\
  \Eprint {http://arxiv.org/abs/1207.0761} {arXiv:1207.0761 [hep-ph]}
  \BibitemShut {NoStop}%
\bibitem [{\citenamefont {Atwood}\ and\ \citenamefont
  {Soni}(2013)}]{Atwood:2012ac}%
  \BibitemOpen
  \bibfield  {author} {\bibinfo {author} {\bibfnamefont {D.}~\bibnamefont
  {Atwood}}\ and\ \bibinfo {author} {\bibfnamefont {A.}~\bibnamefont {Soni}},\
  }\href {\doibase 10.1093/ptep/ptt065} {\bibfield  {journal} {\bibinfo
  {journal} {PTEP}\ }\textbf {\bibinfo {volume} {2013}},\ \bibinfo {pages}
  {0903B05} (\bibinfo {year} {2013})},\ \Eprint
  {http://arxiv.org/abs/1211.1026} {arXiv:1211.1026 [hep-ph]} \BibitemShut
  {NoStop}%
\bibitem [{\citenamefont {Grossman}\ and\ \citenamefont
  {Robinson}(2013)}]{Grossman:2012ry}%
  \BibitemOpen
  \bibfield  {author} {\bibinfo {author} {\bibfnamefont {Y.}~\bibnamefont
  {Grossman}}\ and\ \bibinfo {author} {\bibfnamefont {D.~J.}\ \bibnamefont
  {Robinson}},\ }\href {\doibase 10.1007/JHEP04(2013)067} {\bibfield  {journal}
  {\bibinfo  {journal} {JHEP}\ }\textbf {\bibinfo {volume} {1304}},\ \bibinfo
  {pages} {067} (\bibinfo {year} {2013})},\ \Eprint
  {http://arxiv.org/abs/1211.3361} {arXiv:1211.3361 [hep-ph]} \BibitemShut
  {NoStop}%
\bibitem [{\citenamefont {Hiller}\ \emph {et~al.}(2012)\citenamefont {Hiller},
  \citenamefont {Hochberg},\ and\ \citenamefont {Nir}}]{Hiller:2012wf}%
  \BibitemOpen
  \bibfield  {author} {\bibinfo {author} {\bibfnamefont {G.}~\bibnamefont
  {Hiller}}, \bibinfo {author} {\bibfnamefont {Y.}~\bibnamefont {Hochberg}}, \
  and\ \bibinfo {author} {\bibfnamefont {Y.}~\bibnamefont {Nir}},\ }\href
  {\doibase 10.1103/PhysRevD.85.116008} {\bibfield  {journal} {\bibinfo
  {journal} {Phys.Rev.}\ }\textbf {\bibinfo {volume} {D85}},\ \bibinfo {pages}
  {116008} (\bibinfo {year} {2012})},\ \Eprint {http://arxiv.org/abs/1204.1046}
  {arXiv:1204.1046 [hep-ph]} \BibitemShut {NoStop}%
\bibitem [{\citenamefont {Buccella}\ \emph {et~al.}(2013)\citenamefont
  {Buccella}, \citenamefont {Lusignoli}, \citenamefont {Pugliese},\ and\
  \citenamefont {Santorelli}}]{Buccella:2013tya}%
  \BibitemOpen
  \bibfield  {author} {\bibinfo {author} {\bibfnamefont {F.}~\bibnamefont
  {Buccella}}, \bibinfo {author} {\bibfnamefont {M.}~\bibnamefont {Lusignoli}},
  \bibinfo {author} {\bibfnamefont {A.}~\bibnamefont {Pugliese}}, \ and\
  \bibinfo {author} {\bibfnamefont {P.}~\bibnamefont {Santorelli}},\ }\href
  {\doibase 10.1103/PhysRevD.88.074011} {\bibfield  {journal} {\bibinfo
  {journal} {Phys.Rev.}\ }\textbf {\bibinfo {volume} {D88}},\ \bibinfo {pages}
  {074011} (\bibinfo {year} {2013})},\ \Eprint {http://arxiv.org/abs/1305.7343}
  {arXiv:1305.7343 [hep-ph]} \BibitemShut {NoStop}%
\bibitem [{\citenamefont {Gronau}(2014)}]{Gronau:2013xba}%
  \BibitemOpen
  \bibfield  {author} {\bibinfo {author} {\bibfnamefont {M.}~\bibnamefont
  {Gronau}},\ }\href {\doibase 10.1016/j.physletb.2014.06.055,
  10.1016/j.physletb.2014.01.035} {\bibfield  {journal} {\bibinfo  {journal}
  {Phys.Lett.}\ }\textbf {\bibinfo {volume} {B730}},\ \bibinfo {pages} {221}
  (\bibinfo {year} {2014})},\ \Eprint {http://arxiv.org/abs/1311.1434}
  {arXiv:1311.1434 [hep-ph]} \BibitemShut {NoStop}%
\bibitem [{\citenamefont {Lenz}(2013)}]{Lenz:2013pwa}%
  \BibitemOpen
  \bibfield  {author} {\bibinfo {author} {\bibfnamefont {A.}~\bibnamefont
  {Lenz}},\ }\href@noop {} {\  (\bibinfo {year} {2013})},\ \Eprint
  {http://arxiv.org/abs/1311.6447} {arXiv:1311.6447 [hep-ph]} \BibitemShut
  {NoStop}%
\bibitem [{\citenamefont {Gronau}(2015)}]{Gronau:2015rda}%
  \BibitemOpen
  \bibfield  {author} {\bibinfo {author} {\bibfnamefont {M.}~\bibnamefont
  {Gronau}},\ }\href {\doibase 10.1103/PhysRevD.91.076007} {\bibfield
  {journal} {\bibinfo  {journal} {Phys. Rev.}\ }\textbf {\bibinfo {volume}
  {D91}},\ \bibinfo {pages} {076007} (\bibinfo {year} {2015})},\ \Eprint
  {http://arxiv.org/abs/1501.03272} {arXiv:1501.03272 [hep-ph]} \BibitemShut
  {NoStop}%
\bibitem [{\citenamefont {{M\"uller}}\ \emph {et~al.}(2015)\citenamefont
  {{M\"uller}}, \citenamefont {Nierste},\ and\ \citenamefont
  {Schacht}}]{Muller:2015lua}%
  \BibitemOpen
  \bibfield  {author} {\bibinfo {author} {\bibfnamefont {S.}~\bibnamefont
  {{M\"uller}}}, \bibinfo {author} {\bibfnamefont {U.}~\bibnamefont {Nierste}},
  \ and\ \bibinfo {author} {\bibfnamefont {S.}~\bibnamefont {Schacht}},\ }\href
  {\doibase 10.1103/PhysRevD.92.014004} {\bibfield  {journal} {\bibinfo
  {journal} {Phys. Rev.}\ }\textbf {\bibinfo {volume} {D92}},\ \bibinfo {pages}
  {014004} (\bibinfo {year} {2015})},\ \Eprint
  {http://arxiv.org/abs/1503.06759} {arXiv:1503.06759 [hep-ph]} \BibitemShut
  {NoStop}%
\bibitem [{\citenamefont {Zeppenfeld}(1981)}]{Zeppenfeld:1980ex}%
  \BibitemOpen
  \bibfield  {author} {\bibinfo {author} {\bibfnamefont {D.}~\bibnamefont
  {Zeppenfeld}},\ }\href {\doibase 10.1007/BF01429835} {\bibfield  {journal}
  {\bibinfo  {journal} {Z.Phys.}\ }\textbf {\bibinfo {volume} {C8}},\ \bibinfo
  {pages} {77} (\bibinfo {year} {1981})}\BibitemShut {NoStop}%
\bibitem [{\citenamefont {Gronau}\ \emph {et~al.}(1995)\citenamefont {Gronau},
  \citenamefont {Hernandez}, \citenamefont {London},\ and\ \citenamefont
  {Rosner}}]{Gronau:1995hm}%
  \BibitemOpen
  \bibfield  {author} {\bibinfo {author} {\bibfnamefont {M.}~\bibnamefont
  {Gronau}}, \bibinfo {author} {\bibfnamefont {O.~F.}\ \bibnamefont
  {Hernandez}}, \bibinfo {author} {\bibfnamefont {D.}~\bibnamefont {London}}, \
  and\ \bibinfo {author} {\bibfnamefont {J.~L.}\ \bibnamefont {Rosner}},\
  }\href {\doibase 10.1103/PhysRevD.52.6356} {\bibfield  {journal} {\bibinfo
  {journal} {Phys.Rev.}\ }\textbf {\bibinfo {volume} {D52}},\ \bibinfo {pages}
  {6356} (\bibinfo {year} {1995})},\ \Eprint
  {http://arxiv.org/abs/hep-ph/9504326} {arXiv:hep-ph/9504326 [hep-ph]}
  \BibitemShut {NoStop}%
\bibitem [{\citenamefont {Grossman}\ \emph {et~al.}(2014)\citenamefont
  {Grossman}, \citenamefont {Ligeti},\ and\ \citenamefont
  {Robinson}}]{Grossman:2013lya}%
  \BibitemOpen
  \bibfield  {author} {\bibinfo {author} {\bibfnamefont {Y.}~\bibnamefont
  {Grossman}}, \bibinfo {author} {\bibfnamefont {Z.}~\bibnamefont {Ligeti}}, \
  and\ \bibinfo {author} {\bibfnamefont {D.~J.}\ \bibnamefont {Robinson}},\
  }\href {\doibase 10.1007/JHEP01(2014)066} {\bibfield  {journal} {\bibinfo
  {journal} {JHEP}\ }\textbf {\bibinfo {volume} {1401}},\ \bibinfo {pages}
  {066} (\bibinfo {year} {2014})},\ \Eprint {http://arxiv.org/abs/1308.4143}
  {arXiv:1308.4143 [hep-ph]} \BibitemShut {NoStop}%
\bibitem [{\citenamefont {Grossman}\ and\ \citenamefont
  {Nir}(2012)}]{Grossman:2011zk}%
  \BibitemOpen
  \bibfield  {author} {\bibinfo {author} {\bibfnamefont {Y.}~\bibnamefont
  {Grossman}}\ and\ \bibinfo {author} {\bibfnamefont {Y.}~\bibnamefont {Nir}},\
  }\href {\doibase 10.1007/JHEP04(2012)002} {\bibfield  {journal} {\bibinfo
  {journal} {JHEP}\ }\textbf {\bibinfo {volume} {1204}},\ \bibinfo {pages}
  {002} (\bibinfo {year} {2012})},\ \Eprint {http://arxiv.org/abs/1110.3790}
  {arXiv:1110.3790 [hep-ph]} \BibitemShut {NoStop}%
\bibitem [{\citenamefont {Amhis}\ \emph {et~al.}(2012)\citenamefont {Amhis}
  \emph {et~al.}}]{Amhis:2012bh}%
  \BibitemOpen
  \bibfield  {author} {\bibinfo {author} {\bibfnamefont {Y.}~\bibnamefont
  {Amhis}} \emph {et~al.} (\bibinfo {collaboration} {Heavy Flavor Averaging
  Group}),\ }\href@noop {} {\  (\bibinfo {year} {2012})},\ \Eprint
  {http://arxiv.org/abs/{1207.1158, and online update 30 June 2014}}
  {arXiv:{1207.1158, and online update 30 June 2014} [hep-ex]} \BibitemShut
  {NoStop}%
\bibitem [{\citenamefont {Aubert}\ \emph {et~al.}(2008)\citenamefont {Aubert}
  \emph {et~al.}}]{Aubert:2007if}%
  \BibitemOpen
  \bibfield  {author} {\bibinfo {author} {\bibfnamefont {B.}~\bibnamefont
  {Aubert}} \emph {et~al.} (\bibinfo {collaboration} {BaBar Collaboration}),\
  }\href {\doibase 10.1103/PhysRevLett.100.061803} {\bibfield  {journal}
  {\bibinfo  {journal} {Phys.Rev.Lett.}\ }\textbf {\bibinfo {volume} {100}},\
  \bibinfo {pages} {061803} (\bibinfo {year} {2008})},\ \Eprint
  {http://arxiv.org/abs/0709.2715} {arXiv:0709.2715 [hep-ex]} \BibitemShut
  {NoStop}%
\bibitem [{\citenamefont {Staric}\ \emph {et~al.}(2008)\citenamefont {Staric}
  \emph {et~al.}}]{Staric:2008rx}%
  \BibitemOpen
  \bibfield  {author} {\bibinfo {author} {\bibfnamefont {M.}~\bibnamefont
  {Staric}} \emph {et~al.} (\bibinfo {collaboration} {Belle Collaboration}),\
  }\href {\doibase 10.1016/j.physletb.2008.10.052} {\bibfield  {journal}
  {\bibinfo  {journal} {Phys.Lett.}\ }\textbf {\bibinfo {volume} {B670}},\
  \bibinfo {pages} {190} (\bibinfo {year} {2008})},\ \Eprint
  {http://arxiv.org/abs/0807.0148} {arXiv:0807.0148 [hep-ex]} \BibitemShut
  {NoStop}%
\bibitem [{\citenamefont {Ko}(2013)}]{Ko:2012px}%
  \BibitemOpen
  \bibfield  {author} {\bibinfo {author} {\bibfnamefont {B.}~\bibnamefont {Ko}}
  (\bibinfo {collaboration} {Belle}),\ }\href@noop {} {\bibfield  {journal}
  {\bibinfo  {journal} {PoS}\ }\textbf {\bibinfo {volume} {ICHEP2012}},\
  \bibinfo {pages} {353} (\bibinfo {year} {2013})},\ \Eprint
  {http://arxiv.org/abs/1212.1975} {arXiv:1212.1975} \BibitemShut {NoStop}%
\bibitem [{\citenamefont {{LHCb Collaboration}}(2013)}]{LHCb:2013dka}%
  \BibitemOpen
  \bibfield  {author} {\bibinfo {author} {\bibnamefont {{LHCb
  Collaboration}}},\ }\href@noop {} {\enquote {\bibinfo {title} {{A search for
  time-integrated CP violation in $D^0\rightarrow K^-K^+$ and $D^0\rightarrow
  \pi^-\pi^+$ decays}},}\ } (\bibinfo {year} {2013}),\ \bibinfo {note}
  {{LHCb-CONF-2013-003}}\BibitemShut {NoStop}%
\bibitem [{\citenamefont {Aaij}\ \emph
  {et~al.}(2013{\natexlab{a}})\citenamefont {Aaij} \emph
  {et~al.}}]{Aaij:2013bra}%
  \BibitemOpen
  \bibfield  {author} {\bibinfo {author} {\bibfnamefont {R.}~\bibnamefont
  {Aaij}} \emph {et~al.} (\bibinfo {collaboration} {LHCb collaboration}),\
  }\href {\doibase 10.1016/j.physletb.2013.04.061} {\bibfield  {journal}
  {\bibinfo  {journal} {Phys.Lett.}\ }\textbf {\bibinfo {volume} {B723}},\
  \bibinfo {pages} {33} (\bibinfo {year} {2013}{\natexlab{a}})},\ \Eprint
  {http://arxiv.org/abs/1303.2614} {arXiv:1303.2614 [hep-ex]} \BibitemShut
  {NoStop}%
\bibitem [{\citenamefont {Aaltonen}\ \emph
  {et~al.}(2012{\natexlab{b}})\citenamefont {Aaltonen} \emph
  {et~al.}}]{Aaltonen:2011se}%
  \BibitemOpen
  \bibfield  {author} {\bibinfo {author} {\bibfnamefont {T.}~\bibnamefont
  {Aaltonen}} \emph {et~al.} (\bibinfo {collaboration} {CDF Collaboration}),\
  }\href {\doibase 10.1103/PhysRevD.85.012009} {\bibfield  {journal} {\bibinfo
  {journal} {Phys.Rev.}\ }\textbf {\bibinfo {volume} {D85}},\ \bibinfo {pages}
  {012009} (\bibinfo {year} {2012}{\natexlab{b}})},\ \Eprint
  {http://arxiv.org/abs/1111.5023} {arXiv:1111.5023 [hep-ex]} \BibitemShut
  {NoStop}%
\bibitem [{\citenamefont {Bonvicini}\ \emph {et~al.}(2001)\citenamefont
  {Bonvicini} \emph {et~al.}}]{Bonvicini:2000qm}%
  \BibitemOpen
  \bibfield  {author} {\bibinfo {author} {\bibfnamefont {G.}~\bibnamefont
  {Bonvicini}} \emph {et~al.} (\bibinfo {collaboration} {CLEO Collaboration}),\
  }\href {\doibase 10.1103/PhysRevD.63.071101} {\bibfield  {journal} {\bibinfo
  {journal} {Phys.Rev.}\ }\textbf {\bibinfo {volume} {D63}},\ \bibinfo {pages}
  {071101} (\bibinfo {year} {2001})},\ \Eprint
  {http://arxiv.org/abs/hep-ex/0012054} {arXiv:hep-ex/0012054 [hep-ex]}
  \BibitemShut {NoStop}%
\bibitem [{\citenamefont {Mendez}\ \emph {et~al.}(2010)\citenamefont {Mendez}
  \emph {et~al.}}]{Mendez:2009aa}%
  \BibitemOpen
  \bibfield  {author} {\bibinfo {author} {\bibfnamefont {H.}~\bibnamefont
  {Mendez}} \emph {et~al.} (\bibinfo {collaboration} {CLEO Collaboration}),\
  }\href {\doibase 10.1103/PhysRevD.81.052013} {\bibfield  {journal} {\bibinfo
  {journal} {Phys.Rev.}\ }\textbf {\bibinfo {volume} {D81}},\ \bibinfo {pages}
  {052013} (\bibinfo {year} {2010})},\ \Eprint {http://arxiv.org/abs/0906.3198}
  {arXiv:0906.3198 [hep-ex]} \BibitemShut {NoStop}%
\bibitem [{\citenamefont {Ko}\ \emph {et~al.}(2013)\citenamefont {Ko} \emph
  {et~al.}}]{Ko:2012uh}%
  \BibitemOpen
  \bibfield  {author} {\bibinfo {author} {\bibfnamefont {B.}~\bibnamefont {Ko}}
  \emph {et~al.} (\bibinfo {collaboration} {Belle Collaboration}),\ }\href
  {\doibase 10.1007/JHEP02(2013)098} {\bibfield  {journal} {\bibinfo  {journal}
  {JHEP}\ }\textbf {\bibinfo {volume} {1302}},\ \bibinfo {pages} {098}
  (\bibinfo {year} {2013})},\ \Eprint {http://arxiv.org/abs/1212.6112}
  {arXiv:1212.6112 [hep-ex]} \BibitemShut {NoStop}%
\bibitem [{\citenamefont {Lees}\ \emph {et~al.}(2013)\citenamefont {Lees} \emph
  {et~al.}}]{Lees:2012jv}%
  \BibitemOpen
  \bibfield  {author} {\bibinfo {author} {\bibfnamefont {J.}~\bibnamefont
  {Lees}} \emph {et~al.} (\bibinfo {collaboration} {BaBar Collaboration}),\
  }\href {\doibase 10.1103/PhysRevD.87.052012} {\bibfield  {journal} {\bibinfo
  {journal} {Phys.Rev.}\ }\textbf {\bibinfo {volume} {D87}},\ \bibinfo {pages}
  {052012} (\bibinfo {year} {2013})},\ \Eprint {http://arxiv.org/abs/1212.3003}
  {arXiv:1212.3003 [hep-ex]} \BibitemShut {NoStop}%
\bibitem [{\citenamefont {Link}\ \emph {et~al.}(2002)\citenamefont {Link} \emph
  {et~al.}}]{Link:2001zj}%
  \BibitemOpen
  \bibfield  {author} {\bibinfo {author} {\bibfnamefont {J.}~\bibnamefont
  {Link}} \emph {et~al.} (\bibinfo {collaboration} {FOCUS Collaboration}),\
  }\href {\doibase 10.1103/PhysRevLett.88.041602} {\bibfield  {journal}
  {\bibinfo  {journal} {Phys.Rev.Lett.}\ }\textbf {\bibinfo {volume} {88}},\
  \bibinfo {pages} {041602} (\bibinfo {year} {2002})},\ \Eprint
  {http://arxiv.org/abs/hep-ex/0109022} {arXiv:hep-ex/0109022 [hep-ex]}
  \BibitemShut {NoStop}%
\bibitem [{\citenamefont {Aaij}\ \emph
  {et~al.}(2014{\natexlab{b}})\citenamefont {Aaij} \emph
  {et~al.}}]{Aaij:2014qec}%
  \BibitemOpen
  \bibfield  {author} {\bibinfo {author} {\bibfnamefont {R.}~\bibnamefont
  {Aaij}} \emph {et~al.} (\bibinfo {collaboration} {LHCb Collaboration}),\
  }\href {\doibase 10.1007/JHEP10(2014)025} {\bibfield  {journal} {\bibinfo
  {journal} {JHEP}\ }\textbf {\bibinfo {volume} {1410}},\ \bibinfo {pages} {25}
  (\bibinfo {year} {2014}{\natexlab{b}})},\ \Eprint
  {http://arxiv.org/abs/1406.2624} {arXiv:1406.2624 [hep-ex]} \BibitemShut
  {NoStop}%
\bibitem [{\citenamefont {Ko}\ \emph {et~al.}(2010)\citenamefont {Ko} \emph
  {et~al.}}]{Ko:2010ng}%
  \BibitemOpen
  \bibfield  {author} {\bibinfo {author} {\bibfnamefont {B.}~\bibnamefont {Ko}}
  \emph {et~al.} (\bibinfo {collaboration} {Belle collaboration}),\ }\href
  {\doibase 10.1103/PhysRevLett.104.181602} {\bibfield  {journal} {\bibinfo
  {journal} {Phys.Rev.Lett.}\ }\textbf {\bibinfo {volume} {104}},\ \bibinfo
  {pages} {181602} (\bibinfo {year} {2010})},\ \Eprint
  {http://arxiv.org/abs/1001.3202} {arXiv:1001.3202 [hep-ex]} \BibitemShut
  {NoStop}%
\bibitem [{\citenamefont {Aaij}\ \emph
  {et~al.}(2013{\natexlab{b}})\citenamefont {Aaij} \emph
  {et~al.}}]{Aaij:2013ula}%
  \BibitemOpen
  \bibfield  {author} {\bibinfo {author} {\bibfnamefont {R.}~\bibnamefont
  {Aaij}} \emph {et~al.} (\bibinfo {collaboration} {LHCb collaboration}),\
  }\href {\doibase 10.1007/JHEP06(2013)112} {\bibfield  {journal} {\bibinfo
  {journal} {JHEP}\ }\textbf {\bibinfo {volume} {1306}},\ \bibinfo {pages}
  {112} (\bibinfo {year} {2013}{\natexlab{b}})},\ \Eprint
  {http://arxiv.org/abs/1303.4906} {arXiv:1303.4906 [hep-ex]} \BibitemShut
  {NoStop}%
\bibitem [{\citenamefont {Cenci}(2012)}]{Cenci:2012ps}%
  \BibitemOpen
  \bibfield  {author} {\bibinfo {author} {\bibfnamefont {R.}~\bibnamefont
  {Cenci}} (\bibinfo {collaboration} {BaBar}),\ }\href@noop {} {\  (\bibinfo
  {year} {2012})},\ \Eprint {http://arxiv.org/abs/1209.0138} {arXiv:1209.0138
  [hep-ex]} \BibitemShut {NoStop}%
\bibitem [{\citenamefont {{G.~Hiller, M.~Jung and
  S.~Schacht}}()}]{Hiller:2014prep}%
  \BibitemOpen
  \bibfield  {author} {\bibinfo {author} {\bibnamefont {{G.~Hiller, M.~Jung and
  S.~Schacht}}},\ }\href@noop {} {}\bibinfo {note} {In preparation, DO-TH
  13/25, QFET-2013/12, TTP13-037}\BibitemShut {NoStop}%
\bibitem [{\citenamefont {{Grossman, Yuval and Kagan, Alexander L. and Zupan,
  Jure}}(2012)}]{Grossman:2012eb}%
  \BibitemOpen
  \bibfield  {author} {\bibinfo {author} {\bibnamefont {{Grossman, Yuval and
  Kagan, Alexander L. and Zupan, Jure}}},\ }\href {\doibase
  10.1103/PhysRevD.85.114036} {\bibfield  {journal} {\bibinfo  {journal}
  {Phys.Rev.}\ }\textbf {\bibinfo {volume} {D85}},\ \bibinfo {pages} {114036}
  (\bibinfo {year} {2012})},\ \Eprint {http://arxiv.org/abs/1204.3557}
  {arXiv:1204.3557 [hep-ph]} \BibitemShut {NoStop}%
\bibitem [{\citenamefont {Wiebusch}(2013)}]{Wiebusch:2012en}%
  \BibitemOpen
  \bibfield  {author} {\bibinfo {author} {\bibfnamefont {M.}~\bibnamefont
  {Wiebusch}},\ }\href {\doibase 10.1016/j.cpc.2013.06.008} {\bibfield
  {journal} {\bibinfo  {journal} {Comput.Phys.Commun.}\ }\textbf {\bibinfo
  {volume} {184}},\ \bibinfo {pages} {2438} (\bibinfo {year} {2013})},\ \Eprint
  {http://arxiv.org/abs/1207.1446} {arXiv:1207.1446 [hep-ph]} \BibitemShut
  {NoStop}%
\end{thebibliography}%

\end{document}